\documentclass[conference, bookmarks=false]{IEEEtran}
\pdfoutput=1
\IEEEoverridecommandlockouts

\usepackage[T1]{fontenc}

\usepackage{orcidlink}
\usepackage{float}
\usepackage{microtype}
\usepackage{lipsum}
\usepackage{amssymb}
\usepackage{todonotes}
\usepackage{graphicx}
\usepackage[most]{tcolorbox}
\usepackage{tubscolors}
\usetikzlibrary{%
	shapes,
	automata,
	arrows,
	matrix,
	backgrounds,
	fit,
	patterns,
	decorations.markings, 
	svg.path, 
	shapes.multipart, 
	shadows, 
	fadings,
	shadings,
	positioning, 
	calc,
	decorations, 
	decorations.text,
	intersections, 
	angles, 
	quotes,
	decorations.pathmorphing, 
	arrows.meta, 
	patterns, 
	decorations.pathreplacing, 
	calligraphy,
	snakes,
	bending,   			 	 
	babel,
	spath3%
}   		 	 
\usetikzlibrary{bending}
\usepackage{tikz-3dplot}
\usepackage{tkz-euclide}
\usepackage{tikz-uml}
\usepackage{pgfplots}
\usepackage{pgfplotstable}
\usepackage{tikz-dimline}
\usepgfplotslibrary{fillbetween}
\usetikzlibrary{ext.paths.ortho}

\pgfdeclarelayer{background}%
\pgfdeclarelayer{pre main}%
\pgfdeclarelayer{foreground}%
\pgfsetlayers{background,connections,pre main,main,foreground}%

\input{fig/tikz/shapes.tex}

\tikzset
{
	set stroke arrow/.code={%
		\pgfqkeys{/tikz/stroke arrow}{#1}},
	set stroke arrow={%
		stroke width/.initial=.1em,
		stroke color/.initial=black,
		arrow tip start/.initial=,
		arrow tip end/.initial=,
		tip length/.initial=\the\pgflinewidth,
		tip width/.initial=1.5*\the\pgflinewidth,
		new tip width/.initial=1,
		new tip length/.initial=1,
		shorten front/.initial=0,
		shorten back/.initial=0
	},
	do calc/.code={%
		\pgfmathsetmacro{\b}{\pgfkeysvalueof{/tikz/stroke arrow/tip width}/2}
		\pgfmathsetmacro{\a}{\pgfkeysvalueof{/tikz/stroke arrow/tip length}}
		\pgfmathsetmacro{\c}{sqrt(\a*\a + \b*\b)}
		\pgfmathsetmacro{\hOld}{\a * \b / \c}
		\pgfmathsetmacro{\newwidth}{2*(\hOld+1.5*\pgfkeysvalueof{/tikz/stroke arrow/stroke width})*\c / \a}
		\pgfmathsetmacro{\newlength}{(\hOld+1.5*\pgfkeysvalueof{/tikz/stroke arrow/stroke width})*\c / \b}
		\pgfkeyssetvalue{/tikz/stroke arrow/new tip width}{\newwidth}
		\pgfkeyssetvalue{/tikz/stroke arrow/new tip length}{\newlength}
	},
	stroke arrow/.search also={/tikz},
	stroke arrow/.style= {
		set stroke arrow={#1},
		do calc,
		shorten <= \pgfkeysvalueof{/tikz/stroke arrow/shorten front},
		shorten >=\pgfkeysvalueof{/tikz/stroke arrow/shorten back},
		{\pgfkeysvalueof{/tikz/stroke arrow/arrow tip start}[width=\pgfkeysvalueof{/tikz/stroke arrow/tip width}, length=\pgfkeysvalueof{/tikz/stroke arrow/tip length}]}-{\pgfkeysvalueof{/tikz/stroke arrow/arrow tip end}[width=\pgfkeysvalueof{/tikz/stroke arrow/tip width}, length=\pgfkeysvalueof{/tikz/stroke arrow/tip length}]},
		preaction={%
			draw=\pgfkeysvalueof{/tikz/stroke arrow/stroke color},
			shorten <=\pgfkeysvalueof{/tikz/stroke arrow/shorten front}-1.6*\pgfkeysvalueof{/tikz/stroke arrow/stroke width},
			shorten >= \pgfkeysvalueof{/tikz/stroke arrow/shorten back}-1.6*\pgfkeysvalueof{/tikz/stroke arrow/stroke width},			
			line width=\the\pgflinewidth + \pgfkeysvalueof{/tikz/stroke arrow/stroke width},
			{\pgfkeysvalueof{/tikz/stroke arrow/arrow tip start}[width=\pgfkeysvalueof{/tikz/stroke arrow/new tip width}, length=\pgfkeysvalueof{/tikz/stroke arrow/new tip length}]}-{\pgfkeysvalueof{/tikz/stroke arrow/arrow tip end}[width=\pgfkeysvalueof{/tikz/stroke arrow/new tip width}, length=\pgfkeysvalueof{/tikz/stroke arrow/new tip length}]},
		},
	},
}

\tikzset
{
	set road/.code={\pgfqkeys{/tikz/road}{#1}},
	set road={%
		marking width/.initial=\the\pgflinewidth,%
		lane width/.initial=5*\the\pgflinewidth,%
		dash length/.initial=\pgfkeysvalueof{/tikz/road/marking width},%
		color/.initial=lightgray,%
		opacity/.initial=1,%
		curb/.initial=0.2*\pgfkeysvalueof{/tikz/road/lane width},%
		curb left/.initial=\pgfkeysvalueof{/tikz/road/curb},%
		curb right/.initial=\pgfkeysvalueof{/tikz/road/curb},%
		proxy marking right/.style={solid, white, line width=\pgfkeysvalueof{/tikz/road/marking width}},
		marking right/.style={proxy marking right/.style={#1}},
		proxy marking left/.style={solid, white, line width=\pgfkeysvalueof{/tikz/road/marking width}},
		marking left/.style={proxy marking left/.style={#1}},
		proxy marking center/.style={white, dash pattern=on \pgfkeysvalueof{/tikz/road/dash length} off \pgfkeysvalueof{/tikz/road/dash length}, line width=\pgfkeysvalueof{/tikz/road/marking width}},
		marking center/.style={proxy marking center/.style={#1}}
	},
	road/.search also={/tikz},
	road/.style= {
		draw=none,
		set road={#1},
		preaction={
			line width=2*\pgfkeysvalueof{/tikz/road/lane width},
			draw=\pgfkeysvalueof{/tikz/road/color},
			opacity=\pgfkeysvalueof{/tikz/road/opacity},
			shorten >= 0.1ex
		},
		postaction={decorate},
		decoration={
			markings,
			mark=at position 1ex with
			{
			},
			mark=between positions 0 and 1 step 0.5ex with
			{
				\pgfkeysgetvalue{/pgf/decoration/mark info/sequence number}{\j}
				\pgfmathtruncatemacro{\i}{\j-1}
				\coordinate (A\i) at (0,\pgfkeysvalueof{/tikz/road/lane width});
				\coordinate (B\i) at (0,0);
				\coordinate (C\i) at (0,-\pgfkeysvalueof{/tikz/road/lane width});
			},
			mark=at position 0.5 with
			{
				\pgfkeysgetvalue{/pgf/decoration/mark info/sequence number}{\j}
				\pgfmathtruncatemacro{\marks}{\j-2}
				\pgfmathparse{\pgfkeysvalueof{/tikz/road/curb right}}
				\ifnum 0=\pgfmathresult
				\else
				\draw[solid,opacity=\pgfkeysvalueof{/tikz/road/opacity},\pgfkeysvalueof{/tikz/road/color}, line width=\pgfkeysvalueof{/tikz/road/curb right}] ($(C\marks.center)+(0,-0.5*\pgfkeysvalueof{/tikz/road/curb right}-0.4*\pgfkeysvalueof{/tikz/road/marking width})$)
				foreach \i in {\marks,...,1}
				{ -- ($(C\i.center)+(0,-0.5*\pgfkeysvalueof{/tikz/road/curb right}-0.4*\pgfkeysvalueof{/tikz/road/marking width})$) } -- ($(C1.center)+(0,-0.5*\pgfkeysvalueof{/tikz/road/curb right}-0.4*\pgfkeysvalueof{/tikz/road/marking width})$);
				\fi
				\pgfmathparse{\pgfkeysvalueof{/tikz/road/curb left}}
				\ifnum 0=\pgfmathresult
				\else	            		
				\draw[solid,opacity=\pgfkeysvalueof{/tikz/road/opacity},\pgfkeysvalueof{/tikz/road/color}, line width=\pgfkeysvalueof{/tikz/road/curb left}] ($(A\marks.center)-(0,-0.5*\pgfkeysvalueof{/tikz/road/curb left}-0.4*\pgfkeysvalueof{/tikz/road/marking width})$)
				foreach \i in {\marks,...,1}
				{ -- ($(A\i.center)-(0,-0.5*\pgfkeysvalueof{/tikz/road/curb left}-0.4*\pgfkeysvalueof{/tikz/road/marking width})$) } -- ($(A1.center)-(0,-0.5*\pgfkeysvalueof{/tikz/road/curb left}-0.4*\pgfkeysvalueof{/tikz/road/marking width})$);
				\fi
				\draw[road/proxy marking right] (C\marks.center)
				foreach \i in {\marks,...,1}
				{ -- (C\i.center) } -- (C1.center);
				\draw[road/proxy marking left] (A\marks.center)
				foreach \i in {\marks,...,1}
				{ -- (A\i.center) } -- (A1.center);
				\draw[road/proxy marking center] (B\marks.center)
				foreach \i in {\marks,...,1}
				{ -- (B\i.center) } -- (B1.center);
			}
		},
	},
}
\makeatletter
\pgfdeclaredecoration{free hand}{start}
{
	\state{start}[width = +0pt,
	next state=step,
	persistent precomputation = \pgfdecoratepathhascornerstrue]{}
	\state{step}[auto end on length    = 3pt,
	auto corner on length = 3pt,               
	width=+2pt]
	{
		\pgfpathlineto{
			\pgfpointadd
			{\pgfpoint{2pt}{0pt}}
			{\pgfpoint{rand*0.15pt}{rand*0.15pt}}
		}
	}
	\state{final}
	{}
}
\makeatother
\tikzset{
	double yellow/.style = {%
		double=lightgray,
		double distance=0.8*\mylinewidth,
		line width=\mylinewidth,
		draw=tuYellow
	}
}
\tikzset{
	dashed marking/.style args={#1 dashlength #2}{%
		#1,
		dash pattern=on #2 off #2
	},
	dashed centerline german/.style 2 args={%
		dashed marking={solid, white, line width=#1 dashlength #2}
	}
}
\tikzset{%
	place nodes/.style={%
		postaction={%
			decorate,%
			decoration={%
				markings,%
				mark=between positions 0 and 1 step 0.125 with {%
					\node [] (a\pgfkeysvalueof{/pgf/decoration/mark info/sequence number}) {};%
				},%
			}%
		}%
	}%
}%

\pgfdeclaredecoration{triangle}{start}{
	\state{start}[width=0.99\pgfdecoratedinputsegmentremainingdistance,next state=up from center]
	{\pgfpathlineto{\pgfpointorigin}}
	\state{up from center}[next state=do nothing]
	{
		\pgfpathlineto{\pgfqpoint{\pgfdecoratedinputsegmentremainingdistance}{\pgfdecorationsegmentamplitude}}
		\pgfpathlineto{\pgfqpoint{\pgfdecoratedinputsegmentremainingdistance}{-\pgfdecorationsegmentamplitude}}
		\pgfpathlineto{\pgfpointdecoratedpathfirst}
	}
	\state{do nothing}[width=\pgfdecorationsegmentlength,next state=do nothing]{
		\pgfpathlineto{\pgfpointdecoratedinputsegmentfirst}
		\pgfpathmoveto{\pgfpointdecoratedinputsegmentlast}
	}
}

\tikzset{%
	stop sign/.style 2 args={%
		rounded corners=#1*0.2pt,%
		xshift=#1*\lanewidth,%
		shape=regular polygon,%
		line width=#1*0.8pt,%
		regular polygon sides=8,%
		fill=tuRed,%
		draw=white,%
		inner sep=#1*-0.2pt,%
		label={%
			[anchor=center, 
			yshift=#1*-0.7em, 
			text=white, 
			font=#2\sffamily\bfseries]
			\scalebox{0.5}[1.2]{STOP}},%
		minimum width=#1*1.4em,%
		append after command={\pgfextra%
			\draw[line width=#1*0.3pt, black, rounded corners=#1*0.1pt]%
			($(\tikzlastnode.center)!1.01!(\tikzlastnode.corner 1)$) --%
			($(\tikzlastnode.center)!1.01!(\tikzlastnode.corner 2)$) --%
			($(\tikzlastnode.center)!1.01!(\tikzlastnode.corner 3)$) --%
			($(\tikzlastnode.center)!1.01!(\tikzlastnode.corner 4)$) --%
			($(\tikzlastnode.center)!1.01!(\tikzlastnode.corner 5)$) --%
			($(\tikzlastnode.center)!1.01!(\tikzlastnode.corner 6)$) --%
			($(\tikzlastnode.center)!1.01!(\tikzlastnode.corner 7)$) --%
			($(\tikzlastnode.center)!1.01!(\tikzlastnode.corner 8)$) -- cycle;%
			\endpgfextra%
		}%
	}%
}
\tikzset{%
	psm fact/.style={%
		ellipse,
		drop shadow,
		shadedGray,
		align=center,
		text width=6em,
		minimum height=3em
	},
	detection fact/.style={
		psm fact,
		shadedBlueLight
	},
	action fact/.style={
		psm fact,
		shadedRedLight
	},	
	rule fact/.style={
		psm fact,
		shadedGreenLight
	}	
}

\tikzset{
	triangle path/.style={decoration={triangle,amplitude=#1}, decorate},
	triangle path/.default=1ex}

\tikzset{%
	capability block/.style={
		minimum width=5.5em,
		minimum height=4.5em,
		align=center,
	},
	indicator/.style={
		capability block,
		shadedGray,
	},
	capability fact/.style={%
		capability block,
		shadedBlueLight,
		rounded corners=0.5em,
	},
	myCapability/.style={%
	trapezium,
	trapezium left angle=80,
	trapezium right angle=-80,
	minimum width=4em,
	minimum height=4.2em,
	text width=7.5em,
	align=center,
	inner sep=0.5em
},
	knowledge cap/.style={%
		myCapability,
		shadedGrayLight,
	},
	scenario cap/.style={%
		myCapability,
		shadedGrayDark,
	},
	abstract cap/.style={%
		scenario cap,
		dashed
	},
	connect/.style={%
		thick,
		-{[round]Stealth},
		rounded corners=2pt
	}	
}

\tikzstyle{shadedRedLight} = [top color=white!90!tuRed, bottom color=tuRed!60!white, draw=tuRed80, thick]%
\tikzstyle{shadedBlue} = [top color=tuDarkBlue20, bottom color=tuDarkBlue40, draw=tuDarkBlue80, thick]%
\tikzstyle{shadedBlueLight} = [top color=tuLightBlue20, bottom color=tuLightBlue40, draw=tuBlue, thick]%
\tikzstyle{shadedRed} = [top color=tuRed20, bottom color=tuRed40, draw=tuRed80, thick]%
\tikzstyle{shadedOrange}  = [top color=tuOrange20, bottom color=tuOrange60,  draw=tuOrange100, thick]%
\tikzstyle{shadedOrangeDark}  = [top color=tuOrange40!99!black, bottom color=tuOrange80!99!black,  draw=tuOrange100!99!black, thick]%
\tikzstyle{shadedOrangeLight}  = [top color=tuOrange20,bottom color=tuOrange40, draw=tuOrange80, thick]%
\tikzstyle{shadedGreen}  = [top color=tuGreen20, bottom color=tuGreen60,  draw=tuGreen100, thick]%
\tikzstyle{shadedGreenLight}  = [top color=tuLightGreen20, bottom color=tuLightGreen60,  draw=tuLightGreen100, thick]%
\tikzstyle{shadedYellow} = [top color=tuYellow20, bottom color=tuYellow60,  draw=tuYellow100, thick]%
\tikzstyle{shadedGray} = [top color=white, bottom color=tuGray20, draw=tuGray60, thick]%
\tikzstyle{shadedGrayLight} = [top color=tuBlack!5, bottom color=tuBlack!10, draw=tuBlack!30, thick]%
\tikzstyle{shadedGrayDark} = [top color=tuBlack!20, bottom color=tuBlack!40, draw=tuBlack!60, thick]%
\tikzstyle{shadedGrayMedium} = [top color=tuBlack!13, bottom color=tuBlack!25, draw=gray, thick]%
\tikzstyle{shadedGrayMediumLight} = [top color=tuBlack!10, bottom color=tuBlack!18, draw=gray, thick]%
\tikzstyle{shadedGrayMediumDark} = [top color=tuBlack!15, bottom color=tuBlack!32, draw=darkgray, thick]%
\tikzstyle{shadedBlueMedium} = [top color=tuLightBlue40, bottom color=tuBlue40, draw=tuBlue, thick]%
\tikzstyle{shadedViolet} = [top color=white!97!tuViolet, bottom color=tuViolet!30!white, draw=tuViolet80, thick]%
\tikzstyle{shadow} = [drop shadow={opacity=.5,shadow xshift=.3ex,shadow yshift=-.3ex}]%
\tikzstyle{triangle} = [isosceles triangle,isosceles triangle stretches]%
\tikzstyle{label-it} = [font=\itshape]%
\tikzstyle{block} = [draw, shadedBlue, rectangle, rounded corners, minimum height=2em, minimum width=5em]%
\tikzstyle{smallblock} = [draw, shadedBlue, rectangle, rounded corners, minimum height=1em, minimum width=2em,shadow]%
\tikzstyle{outerblock} = [draw, shadedGrayLight, draw=tuGray80, rectangle, rounded corners, minimum height=2em, minimum width=5em,shadow]%
\tikzstyle{bubble} = [fill=black,shadow,circle,draw=black,inner sep=0pt,minimum size=5pt]%
\tikzstyle{memory} = [cylinder, shape border rotate=90, aspect=.4, shadedGrayLight, minimum width=5em, shadow]%
\tikzstyle{inheritArrow} = [-open triangle 60,thick]%
\tikzstyle{kompArrow}    = [diamond-,thick]%
\tikzstyle{flowDecision} = [diamond, draw, shadedRed, text badly centered, inner sep=0pt,shadow]%
\tikzstyle{flowBlock} = [rectangle, draw, shadedBlue, text centered, rounded corners, minimum height=2em,shadow]%
\tikzstyle{bgBox} = [rectangle, draw, shadedGrayLight, text centered, rounded corners=3mm, shadow, inner sep=10pt]%
\tikzstyle{blockarrow} = [draw, thick, single arrow, minimum height=3em]%

\tikzstyle {archblock} = [outerblock, minimum height=4em, align=center, minimum width=12em, font=\sffamily]
\tikzstyle {slim} = [minimum width=6em]
\tikzstyle {verticalblock} = [archblock, rotate = 90, archblock, minimum width=16.5em, minimum height=2em, on grid]
\tikzstyle {sensors} = [rectangle split parts=2, rectangle split horizontal]

\tikzstyle {sup} = [yshift=0.5cm]
\tikzstyle {sdown} = [yshift=-0.5cm]
\tikzstyle {sleft} = [xshift=-0.5cm]
\tikzstyle {sright} = [xshift=0.5cm]

\tikzstyle {arrow} = [very thick, -{Stealth[round]}]

\tikzstyle {seperator} = [thick, dashed, lightgray]
\tikzstyle {arrow} = [-stealth', thick]
\tikzstyle {rarrow} = [arrow, stealth'-]

\tikzstyle{maneuver area} = [fill=tuDarkBlue40, fill opacity=0.6]
\tikzstyle{blue car} = [vehicle, draw=darkgray!20!black, fill=tuBlue]
\tikzstyle{red car} = [vehicle, draw=darkgray!20!black, fill=tuRed100]
\tikzstyle{orange car} = [vehicle, draw=darkgray!20!black, fill=tuOrange]
\tikzstyle{yellow car} = [vehicle, draw=darkgray!20!black, fill=tuYellow]
\tikzstyle{green car} = [vehicle, draw=darkgray!20!black, fill=tuGreen]
\tikzstyle{target pose} = [opacity=0.7, draw=white]

\makeatletter
\tikzset{reset preactions/.code={\def\tikz@preactions{}}}
\makeatother

\tikzstyle {disable} = [reset preactions, draw=none, fill=none, top color=white, bottom color=white]
\tikzstyle {header} = [disable, font=\sffamily]

\tikzset{
	fit label/.style={yshift={(height("#1")+4pt)/2},
	inner ysep={(height("#1")+8pt)/2},
	label={[anchor=north,font=\itshape]north:#1 \pgfkeysvalueof{/pgf/inner xsep}}},
	free hand/.style={
			decorate, thick, draw,
			decoration={free hand}
	},  
}

\tikzset{
	double arrow/.style args={#1 #2 colored by #3 and #4}{
		-{#1[round]},line width=#2,#3, 
		postaction={draw,-{#1[round, scale=1.095]},#4,line width=(#2)/2,
			shorten <=0.6*(#2)/2,shorten >=0.6*(#2)/2}, 
	}
}

\tikzset{mysplit/.style={rectangle split, rectangle split parts=2, rectangle split draw splits=false, inner sep=2.2ex,
		rectangle split horizontal,minimum width=5.5em},
	textstyle/.style={text height=1.5ex,text depth=.25ex}}

\tikzumlset{class width=18ex}

\tikzumlset{draw element=gray!95!black}
\tikzumlset{font=\small}
\tikzstyle{tikzuml simpleclass style}=[shadedGray, shadow, minimum width=5em, align=center, font=\textbf]
\pgfkeys{%
	/tikzuml/relation/style/.style={#1, font=\small}
}

\tikzset{
	TLeft/.style={label={[anchor=east, xshift=1ex]left:$\blacktriangleleft$}},
	TRight/.style={label={[anchor=west, xshift=-1ex]right:\strut$\blacktriangleright$}},
	TUp/.style={label={[anchor=south, yshift=-1ex]above:$\blacktriangle$}},
        TUpR/.style={label={[anchor=west, xshift=-1.5ex]right:$\blacktriangle$}},
        TUpL/.style={label={[anchor=east, xshift=1.5ex]left:$\blacktriangle$}},
	TDown/.style={label={[anchor=north, yshift=1ex]below:$\blacktriangledown$}},
        TDownR/.style={label={[anchor=west, xshift=-1.5ex]right:$\blacktriangledown$}},
        TDownL/.style={label={[anchor=east, xshift=1.5ex]left:$\blacktriangledown$}},
	closer pos/.style={},
	VP/.style={{archblock, font=\normalfont}}
}

\pgfdeclarepatternformonly{swnestripes}{\pgfpoint{0cm}{0cm}}{\pgfpoint{1cm}{1cm}}{\pgfpoint{1cm}{1cm}}
{
	\foreach \i in {0.1, 0.3,...,0.9}
	{
		\pgfpathmoveto{\pgfpoint{\i cm}{0cm}}
		\pgfpathlineto{\pgfpoint{1cm}{1cm - \i cm}}
		\pgfpathlineto{\pgfpoint{1cm}{1cm - \i cm + 0.1cm}}
		\pgfpathlineto{\pgfpoint{\i cm- 0.1cm}{0cm}}
		\pgfpathclose%
		\pgfusepath{fill}
		\pgfpathmoveto{\pgfpoint{0cm}{\i cm}}
		\pgfpathlineto{\pgfpoint{1cm - \i cm}{1cm}}
		\pgfpathlineto{\pgfpoint{1cm - \i cm- 0.1cm}{1cm}}
		\pgfpathlineto{\pgfpoint{0cm}{\i cm+ 0.1cm}}
		\pgfpathclose%
		\pgfusepath{fill}
	}
}
\pgfdeclarepatternformonly{swneStripes}{\pgfpoint{0cm}{0cm}}{\pgfpoint{1cm}{1cm}}{\pgfpoint{1cm}{1cm}}
{
	\foreach \i in {0.1, 0.3,...,0.9}
	{
		\pgfpathmoveto{\pgfpoint{\i cm}{0cm}}
		\pgfpathlineto{\pgfpoint{1cm}{1cm - \i cm}}
		\pgfpathlineto{\pgfpoint{1cm}{1cm - \i cm - 0.1cm}}
		\pgfpathlineto{\pgfpoint{\i cm+ 0.1cm}{0cm}}
		\pgfpathclose%
		\pgfusepath{fill}
		\pgfpathmoveto{\pgfpoint{0cm}{\i cm}}
		\pgfpathlineto{\pgfpoint{1cm - \i cm}{1cm}}
		\pgfpathlineto{\pgfpoint{1cm - \i cm+ 0.1cm}{1cm}}
		\pgfpathlineto{\pgfpoint{0cm}{\i cm- 0.1cm}}
		\pgfpathclose%
		\pgfusepath{fill}
	}
}
\pgfdeclarepatternformonly{senwstripes}{\pgfpoint{0cm}{0cm}}{\pgfpoint{1cm}{1cm}}{\pgfpoint{1cm}{1cm}}
{
	\foreach \i in {0.1, 0.3,...,0.9}
	{
		\pgfpathmoveto{\pgfpoint{0cm}{\i cm}}
		\pgfpathlineto{\pgfpoint{0cm}{\i cm+ 0.1cm}}
		\pgfpathlineto{\pgfpoint{\i cm+ 0.1cm}{0cm}}
		\pgfpathlineto{\pgfpoint{\i cm}{0cm}}
		\pgfpathclose%
		\pgfusepath{fill}
		\pgfpathmoveto{\pgfpoint{1cm}{\i cm}}
		\pgfpathlineto{\pgfpoint{\i cm}{1cm}}
		\pgfpathlineto{\pgfpoint{\i cm + 0.1cm}{1cm}}
		\pgfpathlineto{\pgfpoint{1cm}{\i cm+ 0.1cm}}
		\pgfpathclose%
		\pgfusepath{fill}
	}
}
\pgfdeclarepatternformonly{senwStripes}{\pgfpoint{0cm}{0cm}}{\pgfpoint{1cm}{1cm}}{\pgfpoint{1cm}{1cm}}
{
	\foreach \i in {0.0, 0.2,...,0.8}
	{
		\pgfpathmoveto{\pgfpoint{\i cm}{0cm}}
		\pgfpathlineto{\pgfpoint{0cm}{\i cm }}
		\pgfpathlineto{\pgfpoint{0cm}{\i cm+ 0.1cm}}
		\pgfpathlineto{\pgfpoint{\i cm+ 0.1cm}{0cm}}
		\pgfpathclose%
		\pgfusepath{fill}
		\pgfpathmoveto{\pgfpoint{1cm}{\i cm}}
		\pgfpathlineto{\pgfpoint{\i cm}{1cm}}
		\pgfpathlineto{\pgfpoint{\i cm+ 0.1cm}{1cm}}
		\pgfpathlineto{\pgfpoint{1cm}{\i cm + 0.1cm}}
		\pgfpathclose%
		\pgfusepath{fill}
	}
}

\newcounter{coordinateindex}

\tikzset{
	put coordinates/.style={
		initialize counter/.code={
			\setcounter{coordinateindex}{0}
		},
		initialize counter,
		decoration={
			show path construction,
			moveto code={
				\stepcounter{coordinateindex}
				\coordinate (#1-\thecoordinateindex) at (\tikzinputsegmentfirst);
			},
			lineto code={
				\stepcounter{coordinateindex}
				\coordinate (#1-\thecoordinateindex) at (\tikzinputsegmentlast);
			},
			curveto code={
				\stepcounter{coordinateindex}
				\coordinate (#1-\thecoordinateindex) at (\tikzinputsegmentlast);
			},
			closepath code={
				\stepcounter{coordinateindex}
				\coordinate (#1-\thecoordinateindex) at (\tikzinputsegmentlast);
			},
		},
		postaction={decorate}
	},
	put coordinates/.default=coordinate
}

\newcommand{\PrepareOcclusion}[2]%
{%
	\path[name path=AB] %
	\foreach \i in {1,...,\thecoordinateindex}{%
		(#1) -- ($(#1)!100cm!(#2-\i)$)%
	};%
	\foreach \i  [count=\j] in {1,...,\thecoordinateindex}%
	{%
		\coordinate (#2-shadow-\i) at (#2-\i);%
	}%
	\pgfresetboundingbox%
	\draw (-20,-20) rectangle (20,20);%
	\path[name path=CBB] (current bounding box.north west) -- (current bounding box.north east) -- (current bounding box.south east) -- (current bounding box.south west) -- cycle;%
	\path[execute at begin node={\global\let\t=\t}, name intersections={name=int-#2, of=AB and CBB, sort by=AB, total=\t}];

	\foreach \i [count=\k from \thecoordinateindex+1] in {\t,...,1}%
	{%
		\def\nodename{#2}
		\global\expandafter\let\csname maxIdx\nodename\endcsname\k
		\coordinate (#2-shadow-\k) at (int-#2-\i);%
		
	}%
	\pgfresetboundingbox%
}%
\newcommand{\DrawOcclusion}[2]%
{%
	\def\maxIdx{\csname maxIdx#1\endcsname}
	\draw[#2](#1-shadow-1)%
	\foreach \x in {2,3,...,\maxIdx}%
	{%
		-- (#1-shadow-\x)%
	} -- cycle;%
}%

\tikzstyle{function} = [shadedGray, thick, font=\bfseries, minimum width=13em,, inner ysep=0.75em, align=center]%
\tikzstyle{port} = [rectangle, shadedGray, minimum width=1.5em, minimum height=1.5em]%
\pgfdeclarearrow{%
	name=Contains,%
	parameters= {\the\pgfarrowlength},%
	setup code={%
		\pgfarrowssettipend{0pt}%
		\pgfarrowssetlineend{-\pgfarrowlength}%
		\pgfarrowlinewidth=\pgflinewidth%
		\pgfarrowssavethe\pgfarrowlength%
	},%
	drawing code={%
		\pgfpathcircle{\pgfpoint{-0.5\pgfarrowlength}{0pt}}{0.5\pgfarrowlength}%
		\pgfpathmoveto{\pgfpoint{-\pgfarrowlength}{0}}%
		\pgfpathlineto{\pgfpoint{0}{0}}%
		\pgfpathmoveto{\pgfpoint{-0.5\pgfarrowlength}{0.5\pgfarrowlength}}%
		\pgfpathlineto{\pgfpoint{-0.5\pgfarrowlength}{-0.5\pgfarrowlength}}%
		\pgfusepathqstroke%
	},%
	defaults = { length = 0.75em }%
}
\tikzset{%
	hw_base/.style={%
		function, %
		label={[anchor=north,#1]\emph{\guillemotleft block\guillemotright}},%
	},%
	hw/.style={%
		hw_base,%
		inner ysep=1.5em, text depth=1.5em%
	},%
	connection/.style args={#1#2}{%
		very thick,%
		postaction={%
			decorate,%
			decoration={%
				markings,%
				mark=at position 0.5 with {\arrow{Triangle[scale=2]}, \node[{#1,align=center, text height=1ex, text depth=0.25ex, font=\small}] {#2};}%
			}%
		},%
	},%
	req/.style args={#1#2#3#4}{%
		label={[anchor=north]\emph{\guillemotleft #1\guillemotright}},%
		align=center,%
		minimum width=7em,
		font=\scriptsize\bfseries,%
		text height=1.5em,%
		append after command={%
			\pgfextra{%
				\path let \p1=($(\tikzlastnode.east) - (\tikzlastnode.west)$) in node[{name=\tikzlastnode-sup, anchor=north, align=left, align=left, minimum width=\x1}] at (\tikzlastnode.south) {Id=``#2''\\Text=``#3''#4};%
				\draw[shadedGray] (\tikzlastnode-sup.north west) -- (\tikzlastnode-sup.north east);%
				\begin{pgfonlayer}{pre main}%
					\node[fit=(\tikzlastnode)(\tikzlastnode-sup), shadedGray,drop shadow, inner sep=0, name=\tikzlastnode-req]{};%
				\end{pgfonlayer}%
			}%
		}%
	},%
	rel label/.style args={#1#2}{%
		postaction={ decorate,%
			decoration={ markings,%
				mark=at position 0.5 with \node[{#1,text height=1ex, text depth=0.25ex, font=\scriptsize}] {#2};%
			}%
		}%
	},%
	inport/.style={%
		port,%
		append after command={%
			\pgfextra{%
				\begin{pgfonlayer}{foreground}%
					\draw[thick,-stealth', shorten <=.4em, shorten >=.4em, gray ] (\tikzlastnode.west) -- (\tikzlastnode.east);%
				\end{pgfonlayer}%
			}%
		}%
	},%
	ioport/.style={%
		port,%
		append after command={%
			\pgfextra{%
				\begin{pgfonlayer}{foreground}%
					\draw[thick,stealth'-stealth', shorten <=.3em, shorten >=.3em, gray ] (\tikzlastnode.west) -- (\tikzlastnode.east);%
				\end{pgfonlayer}%
			}%
		}%
	},%
	outport/.style={%
		port,%
		append after command={%
			\pgfextra{%
				\begin{pgfonlayer}{foreground}%
					\draw[thick,-stealth', shorten <=.3em, shorten >=.3em, gray ] (\tikzlastnode.east) -- (\tikzlastnode.west);%
				\end{pgfonlayer}%
			}%
		}%
	},%
	hw_top_port/.style={%
		hw_base={yshift=-1em},%
		text height=2.75em,%
		inner ysep=0.5em,%
	},%
	derived-req/.style={%
		-{Straight Barb[length=0.5em,width=0.7em]},%
		dashed,%
		thick%
	},%
	contained-req/.style={%
		thick,%
		-Contains%
	},%
	association/.style ={%
		draw,
		thick
	},%
	usecase/.style={
		ellipse,
		shadedGray,
		drop shadow,
		inner sep=1em,
		align=center,
		font=\bfseries
	},%
	diagramStyle/.style={%
		shadedGray, drop shadow, inner ysep=2em, yshift=1em, inner xsep=0.5em
	},
	diagramFrame/.style={%
		thick,gray,fill=none
	}
}%

\tikzset{%
	set class/.code={\pgfqkeys{/tikz/class}{#1}},%
	set class={%
		stereotype/.store in=\myStereoType,%
		class name/.store in=\myClassName,%
		parts/.store in=\myPartNum,%
		stereotype=,%
		class name=myClass,%
		parts=1%
	},%
	class/.search also={/tikz},%
	class/.style={%
		set class={#1},%
		align=center,%
		inner ysep=0.25em,%
		font=\bfseries,%
		shadedGray,drop shadow,%
		node contents={\myClassName},%
		minimum width=10em,
		class/add stereotype/.expand once=\myStereoType%
	},%
	class/.cd,
	add mystereotype/.code={%
		\def\argone{#1}
		\ifx\argone\empty
		\pgfkeysalso{minimum height=2.5em}
		\else
		\pgfkeysalso{
		label={[anchor=north]\emph{\guillemotleft\vphantom{ghA}{#1}\guillemotright}},%
		text width={max(6em, 1em + max(width(\string"#1\string"),width(\string"\myClassName\string"))))},%
		text height=2.2em,%
		text depth=0.25em,%
		}
		\fi
	},
	add stereotype/.style={%
		class/add mystereotype={#1}
	}
}

\tikzset{
	cog/.style={
		draw,
		thick,
		minimum size=0.3cm,
		centerofmass,
		cog rotate=#1
	},
	cog/.default=0,
	frame-arrow/.style={thick, -stealth'},
	vehicle-velocity/.style={frame-arrow, tuBlue},
	s-velocity/.style={frame-arrow, tuOrange},
	wheelforce/.style={frame-arrow, tuRed}
}
\pgfkeyssetvalue{/tikz/cog rotate}{0}%
\tikzset{sFrameInvisible/.style={opacity=0}}%

\tikzset{wheelFrameInvisible/.style={opacity=1}}

\tikzset{trajectoryInvisible/.style={opacity=1}}

\def\centerarc[#1](#2)(#3:#4:#5)
{ \draw[#1] ($(#2)+({#5*cos(#3)},{#5*sin(#3)})$) arc (#3:#4:#5); }

\def\labeledcenterarc[#1](#2)(#3,#4)(#5:#6:#7)
{ \draw[#1] ($(#2)+({#7*cos(#5)},{#7*sin(#5)})$) arc (#5:#6:#7) node[midway,#4] {#3}; }    

\tikzset{wheel/.style={draw=gray, thick, rounded corners=.1em, fill=lightgray!50}}

\makeatletter
\tikzset{%
	reverse path/.style={
		decoration={show path construction,
			reverse path,
			moveto code={\pgfpathmoveto{\pgf@decorate@inputsegment@first}},
			lineto code={\pgfpathlineto{\pgf@decorate@inputsegment@last}},
			curveto code={\pgfpathcurveto{\pgf@decorate@inputsegment@supporta}%
				{\pgf@decorate@inputsegment@supportb}{\pgf@decorate@inputsegment@last}},
			closepath code={\pgfpathclose}
		},
		decorate
	}
}
\makeatother

\tikzset{%
	generalization/.append style={%
		reverse path
	},
	specialization/.append style={%
	reverse path
},	
}

\tikzset{%
	set association/.code={\pgfqkeys{/tikz/association}{#1}},%
	set association={%
		pos1/.store in=\fromPos,%
		pos2/.store in=\toPos,%
		mult1/.store in=\fromMult,%
		mult2/.store in=\toMult,%
		arg1/.store in=\fromArg,%
		arg2/.store in=\toArg,%
		argpos1/.store in=\fromArgPos,%
		argpos2/.store in=\toArgPos,%
		multpos1/.store in=\fromMultPos,%
		multpos2/.store in=\toMultPos,%
		stereotype/.store in=\stereotype,%
		pos/.store in=\sPos,
		pos1=1em,%
		pos2=-1.5em,%
		mult1=,%
		mult2=,%
		arg1=,%
		arg2=,%
		argpos1=,
		argpos2=,
		multpos1=,
		multpos2=,
		stereotype=,%
		pos=0.5%
	},%
	association/.search also={/tikz},%
	association/.style={%
		set association={#1},%
		draw,%
		thick,%
		postaction={decorate},%
		decoration={%
			markings,%
			mark=at position \fromPos-0.02 with%
			{%
				\node (A) {};%
				\pgfgetlastxy{\XCoord}{\YCoord}%
				\global\let\oldX\XCoord%
				\global\let\oldY\YCoord%
			},%
			mark=at position \fromPos with%
			{%
				\node (B){};	%
				\pgfmathsetmacro\lenArg{width("\fromArg")}				
				\pgfmathsetmacro\lenMult{width("\fromMult")}
				\pgfgetlastxy{\xlast}{\ylast}%
				\pgfmathsetmacro{\xDiff}{(abs(\xlast)-abs(\oldX))*100}%
				\pgfmathsetmacro{\yDiff}{(abs(\ylast)-abs(\oldY))*100}%
				\pgfmathsetmacro{\intXDiff}{int((\xlast-\oldX)*100)}
				\pgfmathsetmacro{\labelDiff}{int((abs(\lenArg)-abs(\lenMult))*100)))}
				\def\placeFromName{left}%
				\def\placeFromMult{right}%
				\def\anchorFromMult{west}%
				\def\anchorFromName{east}%
				\pgfmathparse{int(abs(\yDiff) - abs(\xDiff))}%
				\pgfmathsetmacro{\hvDiff}{\pgfmathresult}
				\ifnum\pgfmathresult<0%
							
				\edef\placeFromName{above}%
				\edef\placeFromMult{below}%
				\ifnum\labelDiff<0%
				\ifnum\intXDiff>0%
				\edef\anchorFromName{north west}%
				\edef\anchorFromMult{south west}%
				\else%
				\edef\anchorFromName{south east}%
				\edef\anchorFromMult{north east}%
				\fi%
				\else%
				\ifnum\intXDiff>0%
				\edef\anchorFromName{north west}%
				\edef\anchorFromMult{south west}%
				\else%
				\edef\anchorFromName{north east}%
				\edef\anchorFromMult{south east}%
				\fi%
				\fi%
				\else%
				\fi%
				\node[\placeFromName, anchor=\anchorFromName] {\fromArg};%
				\node[\placeFromMult, anchor=\anchorFromMult]{\vphantom{\fromArg}\fromMult};%
			},%
			mark=at position \sPos with
			{	
				\node[above, transform shape] {\stereotype};								
			},%
			mark=at position \toPos-0.02 with
			{
				\node (D){};	
				\pgfgetlastxy{\xlast}{\ylast}
				\global\let\oldX\xlast
				\global\let\oldY\ylast				
			},
			mark=at position \toPos with
			{
				\node(E){};
				\pgfmathsetmacro\lenArg{width("\toArg")}				
				\pgfmathsetmacro\lenMult{width("\toMult")}
				\pgfgetlastxy{\xlast}{\ylast}%
				\pgfmathsetmacro{\xDiff}{(abs(\xlast)-abs(\oldX))*100}%
				\pgfmathsetmacro{\yDiff}{(abs(\ylast)-abs(\oldY))*100}%
				\pgfmathsetmacro{\intXDiff}{int((\xlast-\oldX)*100)}
				\pgfmathsetmacro{\labelDiff}{int((\lenArg-\lenMult)*100)))}
				\def\placeToName{left}%
				\def\placeToMult{right}%

				\pgfmathparse{int(abs(\yDiff) - abs(\xDiff))}%
				\ifnum\pgfmathresult<0%
				\edef\placeToName{above}%
				\edef\placeToMult{below}%
				\ifnum\labelDiff<0%
				\ifnum\intXDiff>0%
				\else%
				\fi%
				\else%
				\ifnum\intXDiff>0%
				\else%
				\fi%
				\fi%
				\else%
				\fi%
				\node[\placeToName] {\toArg};				
				\node[\placeToMult]{\vphantom{\toArg}\toMult};
			},%
		},%
	},%
}%
\tikzset{%
	rel label/.style args={#1#2#3}{%
		postaction={ decorate,%
			decoration={ markings,%
				mark=at position #1 with \node[{#2,text height=1.5ex, text depth=0.25ex}] {#3};%
			}%
		}%
	},%
	TLeft/.style={label={[anchor=east, xshift=1ex]left:$\blacktriangleleft$}},%
	TRight/.style={label={[anchor=west, xshift=-1ex]right:\strut$\blacktriangleright$}},%
	TUp/.style={label={[anchor=south, yshift=-1ex]above:$\blacktriangle$}},%
	TDown/.style={label={[anchor=north, yshift=1ex]below:$\blacktriangledown$}},%
	closer pos/.style={},%
	class/.append style={minimum width=6.5em, inner ysep=0.5em, inner xsep=0.1em, minimum height=2.75em, font=\bfseries\footnotesize},%
	assoc label/.style={align=center, font=\scriptsize},%
	aggregation/.style={%
		association={#1},%
		-{Diamond[open, length=1.2em]}%
	},%
	generalization/.style={
		draw,
		-{Triangle[open, length=0.7em, width=0.8em]}%
	},%
	specialization/.style={
		draw,
		-{Triangle[length=0.7em, width=0.8em]}%
	},%
	composition/.style={%
		association={#1},
		-{Diamond[length=1.3em]}
	},
	part composite/.style={%
		association={#1},
		{Straight Barb}-{Diamond[length=1.1em]}
	},
	part reference/.style={%
		association={#1},
		{Straight Barb}-{Diamond[open,length=1.1em]}
	}
}%
\newcounter{samplingidx}%
\newcounter{coordindex}%
\newcounter{constraintidx}%
\tikzset{%
	stepcounter-coord/.code={%
		\stepcounter{coordindex}%
	},%
	stepcounter-sample/.code={%
		\stepcounter{samplingidx}%
	},%
	stepcounter-constraint/.code={%
		\stepcounter{constraintidx}%
	}%
}%
\tikzset{%
	polyline/.style={%
		decoration={%
			markings,%
			mark=between positions 0 and 1 step 0.05 with {%
				\coordinate[stepcounter-coord] (A\thecoordindex) at (0,0);%
			}%
		},%
		postaction=decorate,%
	}%
}%
\tikzset{%
	tangent/.style={%
		decoration={%
			markings,
			mark=%
			at position #1%
			with%
			{%
				\coordinate (tangent point-\pgfkeysvalueof{/pgf/decoration/mark info/sequence number}) at (0pt,0pt);%
				\coordinate (tangent unit vector-\pgfkeysvalueof{/pgf/decoration/mark info/sequence number}) at (1,0pt);%
				\coordinate (tangent orthogonal unit vector-\pgfkeysvalueof{/pgf/decoration/mark info/sequence number}) at (0pt,1);%
			}%
		},%
		postaction=decorate%
	},%
	use tangent/.style={%
		shift=(tangent point-#1),%
		x=(tangent unit vector-#1),%
		y=(tangent orthogonal unit vector-#1)%
	},%
	use tangent/.default=1%
}%
\tikzset{%
	constraint line/.style={%
		decoration={%
			markings,%
			mark=between positions 0 and 0.85 step 0.05 with {%
				\node[stepcounter-constraint, circle, inner sep=1pt, fill=tuRed] (lwr-c-\theconstraintidx) at (0,\lowerconstraints[\theconstraintidx]) {};%
				\draw[tuLightGreen](0,0) -- (lwr-c-\theconstraintidx);%
				\draw[tuLightGreen] (0,0) -- (0,3);%
			}%
		},%
		postaction=decorate,%
	}%
}%
\tikzset{%
	sampling line/.style={%
		decoration={%
			markings,
			mark=between positions 0 and 1 step 0.05 with {%
				\coordinate[stepcounter-sample] (B\thesamplingidx-left) at (0,3);
				\coordinate[] (B\thesamplingidx-right) at (0,-6);
			}			
		},
		postaction=decorate,
	}
}
\tikzset{cross/.style={cross out, draw=black, minimum size=2*(#1-\pgflinewidth), inner sep=0pt, outer sep=0pt},	cross/.default={1pt}}

\pgfdeclarelayer{background}%
\pgfdeclarelayer{pre main}%
\pgfdeclarelayer{foreground}%
\pgfdeclarelayer{lifelines}%
\pgfdeclarelayer{activity}%
\pgfdeclarelayer{connections}%
\pgfsetlayers{background,lifelines,activity, connections,pre main,main,foreground}%

\def\mylinewidth{50pt}

\usepackage{cleveref}
\usepackage{csquotes}
\usepackage{subcaption}
\usepackage{tabularx}
\usepackage{makecell}
\usepackage{colortbl}
\usepackage{booktabs}
\usepackage{ragged2e}
\usepackage{nicematrix}
\usepackage{color}
\usepackage{hologo}
\usepackage{listings}

\usepackage[
			bibstyle=ieee,
			citestyle=numeric-comp,
			backend=biber,
			minnames=1,
			maxcitenames=2,
			maxbibnames=99,
			doi=true,
			isbn=false,
			url=false,
			hyperref=false,
			natbib=true]
			{biblatex}

\def\BibTeX{{\rm B\kern-.05em{\sc i\kern-.025em b}\kern-.08em
    T\kern-.1667em\lower.7ex\hbox{E}\kern-.125emX}}

\crefformat{figure}{Fig.~#2#1#3}
\newsavebox{\largestimage}%
\usepackage{xpatch}
\usepackage{xstring}

\DeclareFieldFormat{titlecase}{\MakeCapital{#1}} \DeclareFieldFormat{sentencecase}{#1}

\DeclareDatamodelEntrytypes{standard}%
\DeclareDatamodelEntryfields[standard]{type,number}%
\DeclareDatamodelFields[type=field,datatype=verbatim]{subtype}%
\DeclareDatamodelEntryfields[report]{subtype}%
\DeclareFieldFormat{subtype}{#1}

\DeclareSourcemap{
	\maps[datatype=bibtex]{
		\map{
			\step[fieldsource=langid, match=\regexp{\A(n)?((swiss)?german|austrian)\Z}, final]
			\step[fieldset=language, fieldvalue={(in German)}]
		}
		\map{
			\step[fieldsource=langid, match=pinyin, final]
			\step[fieldset=language, fieldvalue={(in Chinese)}]
		}
		\map{
			\step[fieldsource=langid, match=japanese, final]
			\step[fieldset=language, fieldvalue={(in Japanese)}]
		}
		\map{
			\step[fieldsource=langid, match=ko, final]
			\step[fieldset=language, fieldvalue={(in Korean)}]
		}
        \map[overwrite]{%
          \pertype{report}                   
          \step[fieldsource=number,match=\regexp{(ISO|BSI|DIN|IWA|DIN|IEC|IEEE)}, final]
          \step[fieldset=keywords, fieldvalue={type-in-number},append]
        }
        \map{%
          \pertype{report}                   
          \step[fieldsource=number,match=\regexp{(ISO|BSI|DIN|IWA|DIN|IEC|IEEE)}]
          \step[fieldset=annotation, fieldvalue={Standard}]
        }        
        \map[overwrite]{%
          \pertype{report}
          \step[fieldsource=number,match=\regexp{TR}, final]
          \step[fieldset=annotation, fieldvalue={Tech. Rep.}]
        }
        \map[overwrite]{%
          \pertype{report}
          \step[fieldsource=number,match=\regexp{PAS}, final]
          \step[fieldset=annotation, fieldvalue={Publ. Avail. Spec.}]
        }
        \map[overwrite]{%
          \pertype{report}
          \step[fieldsource=number,match=\regexp{TS}, final]
          \step[fieldset=annotation, fieldvalue={Tech. Spec.}]
        }
        \map[overwrite]{%
          \pertype{report}
          \step[fieldsource=number,match=\regexp{IWA}, final]
          \step[fieldset=annotation, fieldvalue={Intern. Workshop Agreement}]
        }
        \map[overwrite]{%
          \pertype{report}
          \step[fieldsource=number,match=\regexp{Guide}, final]
          \step[fieldset=annotation, fieldvalue={Guide}]
        }
        \map[overwrite]{%
          \pertype{report}
          \step[fieldsource=number, match=\regexp{.*\s+((?!v)\S+)$}, final]
          \step[fieldset=number, fieldvalue={$1}]
        }        
    }
    \maps[datatype=biber]{
		\map{
			\step[fieldsource=note, final]
			\step[fieldset=addendum, origfieldval, final]
			\step[fieldset=note, null]
	  }
    }
}

\xpatchbibdriver{inproceedings}					 	
	{\usebibmacro{publisher+location+date}}			
	{\IfSubStr{\strfield{booktitle}}{\strfield{year}}{\usebibmacro{publisher+location}}{\usebibmacro{publisher+location+date}}} 
	{} 
	{\typeout{There was an error patching biblatex-ieee (specifically, ieee.bbx's @inproceedings driver)}} 

\newbibmacro*{publisher+location}{%
	\printlist{location}%
	\iflistundef{publisher}
	{\setunit*{\addcomma\space}}
	{\setunit*{\addcolon\space}}%
	\printlist{publisher}%
	\newunit}

\xpatchbibdriver{online}
{\printtext[parens]{\usebibmacro{date}}}
	{\iffieldundef{year}
		{}
		{\printtext[parens]{\usebibmacro{date}}}}
	{}
	{\typeout{There was an error patching biblatex-ieee (specifically, ieee.bbx's @online driver)}}

\DeclareBibliographyDriver{standard}{%
        \printfield{title}
	\printfield{shorttitle}\addcolon%
	\printfield{year}%
	\addspace
	\parentext{\printfield{year}}
	\setunit{\labelnamepunct}\newblock
	\usebibmacro{title}%
	\newunit\newblock
	\printnames{author}%
	\setunit{\addspace}\newblock
	\printfield[parens]{type}%
	\newunit\newblock
	\iftoggle{bbx:url}
	{\usebibmacro{url+urldate}}
	{}%
	\newunit\newblock
	\usebibmacro{addendum+pubstate}%
	\setunit{\bibpagerefpunct}\newblock
	\usebibmacro{pageref}%
	\newunit\newblock
	\usebibmacro{related}%
	\usebibmacro{finentry}}

\DeclareFieldFormat[report]{titlecase}{#1}

\DeclareFieldFormat[report]{title}{%
    \iffieldequalstr{type}{standard}{%
       \mkbibemph{#1}%
    }{%
        \mkbibquote{#1}%
    }%
}%
\DeclareFieldFormat[report]{type}{\MakeCapital{#1}}

\newbibmacro*{standardno}{%
    \printlist{institution}%
    \setunit{\addspace}%
    \iffieldundef{note}{%
        \printfield{annotation}%
    }{%
        \printfield{note}%
    }%
    \setunit{\addspace}
    \printfield{number}%
}

\DeclareBibliographyDriver{report}{%
  \iffieldequalstr{type}{standard}{%
          \usebibmacro{bibindex}%
          \usebibmacro{begentry}%
          \printfield{title}%
          \setunit{\labelnamepunct}\newblock
          \usebibmacro{standardno}%
          \addcolon
          \usebibmacro{date}
  }{%
      \usebibmacro{bibindex}%
      \usebibmacro{begentry}%
      \usebibmacro{author}%
      \setunit{\labelnamepunct}\newblock
      \usebibmacro{title}%
      \newunit
      \printlist{language}%
      \newunit\newblock
      \usebibmacro{byauthor}%
      \newunit\newblock
      \usebibmacro{institution+location}%
      \newunit\newblock
      \printfield{type}%
      \setunit*{\addspace}%
      \printfield{number}%
      \newunit\newblock
      \printfield{version}%
      \newunit\newblock
      \usebibmacro{date}%
      \newunit
      \printfield{note}%
      \newunit\newblock
      \usebibmacro{chapter+pages}%
      \newunit
      \printfield{pagetotal}%
      \newunit\newblock
      \iftoggle{bbx:isbn}
        {\printfield{isrn}}
        {}%
      \newunit\newblock
      \usebibmacro{doi+eprint+url}%
      \newunit\newblock
      \usebibmacro{addendum+pubstate}%
      \setunit{\bibpagerefpunct}\newblock
      \usebibmacro{pageref}%
      \newunit\newblock
      \iftoggle{bbx:related}
        {\usebibmacro{related:init}%
         \usebibmacro{related}}
        {}%
    }
    \usebibmacro{finentry}}

\newbibmacro*{publisher+type+number}{%
	\printtext{%
		\printlist{publisher}
		\iflistundef{publisher}
		{\space}{}%
		\iffieldundef{type}
		{Standard}
		{\printfield{type}}
		\printfield{number}
	}%
}

\xpatchbibdriver{report}					 	
	{\usebibmacro{date}}			
	{\IfSubStr{\strfield{number}}{\strfield{year}}{}{\usebibmacro{date}}} 
	{} 
	{\typeout{There was an error patching biblatex-ieee (specifically, ieee.bbx's @report driver)}} 

\DeclareBibliographyDriver{software}{%
	\usebibmacro{bibindex}%
	\usebibmacro{begentry}%
	\usebibmacro{maintitle+title}%
	\setunit{\addcomma\addspace}%
	\usebibmacro{version+year}%
	\setunit{\labelnamepunct}\newblock
	\usebibmacro{author}%
	\setunit{\addcomma\addspace}%
	\newunit\newblock
	\usebibmacro{url+urldate}%
	\newunit\newblock
	\usebibmacro{finentry}%
}

\newbibmacro*{version+year}{%
	\printtext{%
		\iffieldundef{version}{\iffieldundef{year}{}{\printfield{year}}}{\printfield{version}}
		\printfield{number}
		\printlist{publisher}
		\iflistundef{publisher}{\space}{}%
	}%
}

\makeatletter
\newcounter{IEEE@bibentries}
\renewcommand\IEEEtriggeratref[1]{%
  \renewbibmacro{finentry}{%
    \stepcounter{IEEE@bibentries}%
    \ifthenelse{\equal{\value{IEEE@bibentries}}{#1}}
    {\finentry\@IEEEtriggercmd}
    {\finentry}%
  }%
}
\makeatother
\DeclareSourcemap{%
    \maps{
            \map[overwrite, foreach={booktitle,journaltitle,eventtitle,series,publisher,institution,type}]{
			\step[fieldsource=\regexp{$MAPLOOP}, match={XXX}, replace={Acad. Serbe Sci. Arts Glas Cl. Sci. Tech.}]
			\step[fieldsource=\regexp{$MAPLOOP}, match={XXX}, replace={Acta Acust.}]
			\step[fieldsource=\regexp{$MAPLOOP}, match={XXX}, replace={Acta Astron. (Poland)}]
			\step[fieldsource=\regexp{$MAPLOOP}, match={XXX}, replace={Acta Astron. Sin. (China)}]
			\step[fieldsource=\regexp{$MAPLOOP}, match={XXX}, replace={Acta Astronaut. (U.K.)}]
			\step[fieldsource=\regexp{$MAPLOOP}, match={XXX}, replace={Acta Astrophys. Sin. (China)}]
			\step[fieldsource=\regexp{$MAPLOOP}, match={XXX}, replace={Acta Autom. Sin.}]
			\step[fieldsource=\regexp{$MAPLOOP}, match={XXX}, replace={Acta Cienc. Indica Math.}]
			\step[fieldsource=\regexp{$MAPLOOP}, match={XXX}, replace={Acta Cienc. Indica Phys.}]
			\step[fieldsource=\regexp{$MAPLOOP}, match={XXX}, replace={Acta Crystallogr. A, Found. Crystallogr.}]
			\step[fieldsource=\regexp{$MAPLOOP}, match={XXX}, replace={Acta Crystallogr. B, Struct. Sci.}]
			\step[fieldsource=\regexp{$MAPLOOP}, match={XXX}, replace={Acta Crystallogr. C, Cryst. Struct. Commun.}]
			\step[fieldsource=\regexp{$MAPLOOP}, match={XXX}, replace={Acta Electron. (France)}]
			\step[fieldsource=\regexp{$MAPLOOP}, match={XXX}, replace={Acta Cyberno}]
			\step[fieldsource=\regexp{$MAPLOOP}, match={XXX}, replace={Acta Electron. Sin. (China)}]
			\step[fieldsource=\regexp{$MAPLOOP}, match={XXX}, replace={Acta Geod. Geophys. Montan. Hung.}]
			\step[fieldsource=\regexp{$MAPLOOP}, match={XXX}, replace={Acta Geophys. Pol.}]
			\step[fieldsource=\regexp{$MAPLOOP}, match={XXX}, replace={Acta Geophys. Sin. (China)}]
			\step[fieldsource=\regexp{$MAPLOOP}, match={XXX}, replace={Acta Geophys. Sin. (USA)}]
			\step[fieldsource=\regexp{$MAPLOOP}, match={XXX}, replace={Acta Metall.}]
			\step[fieldsource=\regexp{$MAPLOOP}, match={XXX}, replace={Acta Mex. Cienc. Tecnol.}]
			\step[fieldsource=\regexp{$MAPLOOP}, match={XXX}, replace={Acta Phys. Hung.}]
			\step[fieldsource=\regexp{$MAPLOOP}, match={XXX}, replace={Acta Phys. Pol. A}]
			\step[fieldsource=\regexp{$MAPLOOP}, match={XXX}, replace={Acta Phys. Pol. B}]
			\step[fieldsource=\regexp{$MAPLOOP}, match={XXX}, replace={Acta Phys. Sin.}]
			\step[fieldsource=\regexp{$MAPLOOP}, match={XXX}, replace={Acta Phys. Slovaca}]
			\step[fieldsource=\regexp{$MAPLOOP}, match={XXX}, replace={Acta Politec. Mex.}]
			\step[fieldsource=\regexp{$MAPLOOP}, match={XXX}, replace={Acta Polytech. Scand. Appl. Phys. Ser.}]
			\step[fieldsource=\regexp{$MAPLOOP}, match={XXX}, replace={Acta Polytech. Scand. Chem. Technol.}]
			\step[fieldsource=\regexp{$MAPLOOP}, match={XXX}, replace={Metall. Ser.}]
			\step[fieldsource=\regexp{$MAPLOOP}, match={XXX}, replace={Acta Polytech. Scand. Electr. Eng. Ser.}]
			\step[fieldsource=\regexp{$MAPLOOP}, match={XXX}, replace={Acta Polytech. Scand. Math. Comput. Sci. Ser.}]
			\step[fieldsource=\regexp{$MAPLOOP}, match={XXX}, replace={Acta Polytech. Scand. Mech. Eng. Ser.}]
			\step[fieldsource=\regexp{$MAPLOOP}, match={XXX}, replace={Acta Seismol. Sin.}]
			\step[fieldsource=\regexp{$MAPLOOP}, match={XXX}, replace={Acta Tech. Acad. Sci. Hung.}]
			\step[fieldsource=\regexp{$MAPLOOP}, match={XXX}, replace={Acta Tech. CSAV}]
			\step[fieldsource=\regexp{$MAPLOOP}, match={XXX}, replace={Acustica}]
			\step[fieldsource=\regexp{$MAPLOOP}, match={XXX}, replace={(AEU) Arch. Elektr. Ubertragung}]
			\step[fieldsource=\regexp{$MAPLOOP}, match={XXX}, replace={Akust. Zh.}]
			\step[fieldsource=\regexp{$MAPLOOP}, match={XXX}, replace={Algorithmica}]
			\step[fieldsource=\regexp{$MAPLOOP}, match={XXX}, replace={Alta Freq.}]
			\step[fieldsource=\regexp{$MAPLOOP}, match={XXX}, replace={An. Acad. Bras. Cienc.}]
			\step[fieldsource=\regexp{$MAPLOOP}, match={XXX}, replace={An. Fis.}]
			\step[fieldsource=\regexp{$MAPLOOP}, match={XXX}, replace={An. Mec. Electr.}]
			\step[fieldsource=\regexp{$MAPLOOP}, match={XXX}, replace={Angew. Inform.}]
			\step[fieldsource=\regexp{$MAPLOOP}, match={XXX}, replace={Ann. Inst. Henri Poincare Phys. Theor.}]
			\step[fieldsource=\regexp{$MAPLOOP}, match={XXX}, replace={Ann. Soc. Sci. Brux. I, Sci. Math. Astron. Phys.}]
			\step[fieldsource=\regexp{$MAPLOOP}, match={XXX}, replace={Arch. Elektr. Uebertrag. (AEU)}] 
			\step[fieldsource=\regexp{$MAPLOOP}, match={XXX}, replace={Arch. Elektron. Uebertrag. Tech.}] 
			\step[fieldsource=\regexp{$MAPLOOP}, match={XXX}, replace={Arch. Elektrotech. (Poland)}]
			\step[fieldsource=\regexp{$MAPLOOP}, match={XXX}, replace={Arch. Elektrotech. (Germany)}]
			\step[fieldsource=\regexp{$MAPLOOP}, match={XXX}, replace={Ark. Fys. Semin. Trondheim}]
			\step[fieldsource=\regexp{$MAPLOOP}, match={XXX}, replace={Astrofizika}]
			\step[fieldsource=\regexp{$MAPLOOP}, match={XXX}, replace={Astron. Nachr. (Germany)}]
			\step[fieldsource=\regexp{$MAPLOOP}, match={XXX}, replace={Astron. Tidsskr.}]
			\step[fieldsource=\regexp{$MAPLOOP}, match={XXX}, replace={Astron. Vestn.}]
			\step[fieldsource=\regexp{$MAPLOOP}, match={XXX}, replace={Astron. Zh.}]
			\step[fieldsource=\regexp{$MAPLOOP}, match={XXX}, replace={Atomwirtsch.-Atomtech.}]
			\step[fieldsource=\regexp{$MAPLOOP}, match={XXX}, replace={Atti Accad. Sci. Ist. Bologna CI. Sci. Fis.}]
			\step[fieldsource=\regexp{$MAPLOOP}, match={XXX}, replace={Rend. XIII}]
			\step[fieldsource=\regexp{$MAPLOOP}, match={XXX}, replace={Atti Accad. Sci. Torino I, CI. Sci. Fis. Math. Nat.}]
			\step[fieldsource=\regexp{$MAPLOOP}, match={XXX}, replace={Autom. Strum.}]
			\step[fieldsource=\regexp{$MAPLOOP}, match={XXX}, replace={Autom. Tech. Prax.}]
			\step[fieldsource=\regexp{$MAPLOOP}, match={XXX}, replace={Automatica}]
			\step[fieldsource=\regexp{$MAPLOOP}, match={XXX}, replace={Automatie}]
			\step[fieldsource=\regexp{$MAPLOOP}, match={XXX}, replace={Automatika}]
			\step[fieldsource=\regexp{$MAPLOOP}, match={XXX}, replace={Automatisierungstechnik}]
			\step[fieldsource=\regexp{$MAPLOOP}, match={XXX}, replace={Automatizace}]
			\step[fieldsource=\regexp{$MAPLOOP}, match={XXX}, replace={Automedica}]
			\step[fieldsource=\regexp{$MAPLOOP}, match={XXX}, replace={Avtom. Telemekh.}]
			\step[fieldsource=\regexp{$MAPLOOP}, match={XXX}, replace={Avtom. Vychisl. Tekh.}]
			\step[fieldsource=\regexp{$MAPLOOP}, match={XXX}, replace={Avtomatika}]
			\step[fieldsource=\regexp{$MAPLOOP}, match={XXX}, replace={Avtometriya}]
			\step[fieldsource=\regexp{$MAPLOOP}, match={ATZelektronik worldwide}, replace={ATZelektron. worldw.}]
			\step[fieldsource=\regexp{$MAPLOOP}, match={XXX}, replace={Ber. Bunsenges. Phys. Chem.}]
			\step[fieldsource=\regexp{$MAPLOOP}, match={XXX}, replace={Biofizika}]
			\step[fieldsource=\regexp{$MAPLOOP}, match={XXX}, replace={Biometrika}]
			\step[fieldsource=\regexp{$MAPLOOP}, match={XXX}, replace={Boll. Geofis. Teor. Appl.}]
			\step[fieldsource=\regexp{$MAPLOOP}, match={XXX}, replace={Bull. Acad. Serbe Sci. Arts cl. Sci. Tech.}]
			\step[fieldsource=\regexp{$MAPLOOP}, match={XXX}, replace={Bull. Annu. Soc. Suisse Chronom. Lab. Suisse.}]
			\step[fieldsource=\regexp{$MAPLOOP}, match={XXX}, replace={Rech. Horlog.}]
			\step[fieldsource=\regexp{$MAPLOOP}, match={XXX}, replace={Bull. Cl. Sci. Acad. R. Belg.}]
			\step[fieldsource=\regexp{$MAPLOOP}, match={XXX}, replace={Bull. Dir. Etud. Rech. A}]
			\step[fieldsource=\regexp{$MAPLOOP}, match={XXX}, replace={Bull. Dir. Etud. Rech. B}]
			\step[fieldsource=\regexp{$MAPLOOP}, match={XXX}, replace={Bull. Dir. Etud. Rech. C}]
			\step[fieldsource=\regexp{$MAPLOOP}, match={XXX}, replace={Bull. Liaison Rech. Inform. Autom.}]
			\step[fieldsource=\regexp{$MAPLOOP}, match={XXX}, replace={Bur. Etud. Autom.}]
			\step[fieldsource=\regexp{$MAPLOOP}, match={XXX}, replace={CFI-Ceram. Forum Int.-Ber. Dtsch. Keram. Ges.}]
			\step[fieldsource=\regexp{$MAPLOOP}, match={XXX}, replace={Chem. Scr.}]
			\step[fieldsource=\regexp{$MAPLOOP}, match={XXX}, replace={Ciel Terre}]
			\step[fieldsource=\regexp{$MAPLOOP}, match={XXX}, replace={Cybernetica}]
			\step[fieldsource=\regexp{$MAPLOOP}, match={XXX}, replace={Deut. Hydrogr. Z.}]
			\step[fieldsource=\regexp{$MAPLOOP}, match={XXX}, replace={Dokl. Akad. Nauk SSSR}]
			\step[fieldsource=\regexp{$MAPLOOP}, match={XXX}, replace={Electroacoustique}]
			\step[fieldsource=\regexp{$MAPLOOP}, match={XXX}, replace={Electrochim. Acta}]
			\step[fieldsource=\regexp{$MAPLOOP}, match={XXX}, replace={Electrochim. Metal.}]
			\step[fieldsource=\regexp{$MAPLOOP}, match={XXX}, replace={Elektor Electron.}]
			\step[fieldsource=\regexp{$MAPLOOP}, match={XXX}, replace={Elektr. Bahnen}]
			\step[fieldsource=\regexp{$MAPLOOP}, match={XXX}, replace={Elektr. Energ.-Tech.}]
			\step[fieldsource=\regexp{$MAPLOOP}, match={XXX}, replace={Elektr. Masch.}]
			\step[fieldsource=\regexp{$MAPLOOP}, match={XXX}, replace={Elektr. Stn.}]
			\step[fieldsource=\regexp{$MAPLOOP}, match={XXX}, replace={Elektrichestvo}]
			\step[fieldsource=\regexp{$MAPLOOP}, match={XXX}, replace={Elektrie}]
			\step[fieldsource=\regexp{$MAPLOOP}, match={XXX}, replace={Elektrizitaetswirtschaft}]
			\step[fieldsource=\regexp{$MAPLOOP}, match={XXX}, replace={Elektro}]
			\step[fieldsource=\regexp{$MAPLOOP}, match={XXX}, replace={Elektro-Anz.}]
			\step[fieldsource=\regexp{$MAPLOOP}, match={XXX}, replace={Elektro-Jahr}]
			\step[fieldsource=\regexp{$MAPLOOP}, match={XXX}, replace={Elektrokhimiya}]
			\step[fieldsource=\regexp{$MAPLOOP}, match={XXX}, replace={Elektron}]
			\step[fieldsource=\regexp{$MAPLOOP}, match={XXX}, replace={Elektron Int.}]
			\step[fieldsource=\regexp{$MAPLOOP}, match={XXX}, replace={Elektron. Entwitkl.}]
			\step[fieldsource=\regexp{$MAPLOOP}, match={XXX}, replace={Elektron. Ind}]
			\step[fieldsource=\regexp{$MAPLOOP}, match={XXX}, replace={Elektron. J.}]
			\step[fieldsource=\regexp{$MAPLOOP}, match={XXX}, replace={Elektron. Prax.}]
			\step[fieldsource=\regexp{$MAPLOOP}, match={XXX}, replace={Elektron. Tekh.}]
			\step[fieldsource=\regexp{$MAPLOOP}, match={XXX}, replace={Elektronica}]
			\step[fieldsource=\regexp{$MAPLOOP}, match={XXX}, replace={Elektronik}]
			\step[fieldsource=\regexp{$MAPLOOP}, match={XXX}, replace={Elektronika}]
			\step[fieldsource=\regexp{$MAPLOOP}, match={XXX}, replace={Elektroniker}]
			\step[fieldsource=\regexp{$MAPLOOP}, match={XXX}, replace={Elektronikschau}]
			\step[fieldsource=\regexp{$MAPLOOP}, match={XXX}, replace={Elektrosvyaz}]
			\step[fieldsource=\regexp{$MAPLOOP}, match={XXX}, replace={Elektrotech. Cas.}]
			\step[fieldsource=\regexp{$MAPLOOP}, match={Elektrotechnik und Informationstechnik}, replace={Elektrotech. Inf. Tech.}]
			\step[fieldsource=\regexp{$MAPLOOP}, match={XXX}, replace={Elektrotech. Obz.}]
			\step[fieldsource=\regexp{$MAPLOOP}, match={XXX}, replace={Elektrotechniek}]
			\step[fieldsource=\regexp{$MAPLOOP}, match={XXX}, replace={Elektrotechnik (Czechoslovakia)}]
			\step[fieldsource=\regexp{$MAPLOOP}, match={XXX}, replace={Elektrotechnik (Switzerland)}]
			\step[fieldsource=\regexp{$MAPLOOP}, match={XXX}, replace={Elektrotechnik (Germany)}]
			\step[fieldsource=\regexp{$MAPLOOP}, match={XXX}, replace={Elektrotechnika}]
			\step[fieldsource=\regexp{$MAPLOOP}, match={XXX}, replace={Elektrotehnika, Zagreb}]
			\step[fieldsource=\regexp{$MAPLOOP}, match={XXX}, replace={Elektrotekhnika}]
			\step[fieldsource=\regexp{$MAPLOOP}, match={XXX}, replace={Elektroteknikeren}]
			\step[fieldsource=\regexp{$MAPLOOP}, match={XXX}, replace={Elektrowaerme Int. B.}]
			\step[fieldsource=\regexp{$MAPLOOP}, match={XXX}, replace={Elettrificazione}]
			\step[fieldsource=\regexp{$MAPLOOP}, match={XXX}, replace={Elettron. Oggi}]
			\step[fieldsource=\regexp{$MAPLOOP}, match={XXX}, replace={Elettron. Telecomun.}]
			\step[fieldsource=\regexp{$MAPLOOP}, match={XXX}, replace={Elettrotecnica}]
			\step[fieldsource=\regexp{$MAPLOOP}, match={XXX}, replace={Elek. Med Aktuell Elektron.}]
			\step[fieldsource=\regexp{$MAPLOOP}, match={XXX}, replace={Elteknik}]
			\step[fieldsource=\regexp{$MAPLOOP}, match={XXX}, replace={Energ. Atomtech.}]
			\step[fieldsource=\regexp{$MAPLOOP}, match={XXX}, replace={Energ. Elettr.}]
			\step[fieldsource=\regexp{$MAPLOOP}, match={XXX}, replace={Energetica}]
			\step[fieldsource=\regexp{$MAPLOOP}, match={XXX}, replace={Energetik}]
			\step[fieldsource=\regexp{$MAPLOOP}, match={XXX}, replace={Energetika}]
			\step[fieldsource=\regexp{$MAPLOOP}, match={XXX}, replace={Energetyka}]
			\step[fieldsource=\regexp{$MAPLOOP}, match={XXX}, replace={Energia Nuclear}]
			\step[fieldsource=\regexp{$MAPLOOP}, match={XXX}, replace={Energie Technik (Germany)}]
			\step[fieldsource=\regexp{$MAPLOOP}, match={XXX}, replace={Energie Technik (Switzerland)}]
			\step[fieldsource=\regexp{$MAPLOOP}, match={XXX}, replace={Entropie}]
			\step[fieldsource=\regexp{$MAPLOOP}, match={XXX}, replace={ETZ}]
			\step[fieldsource=\regexp{$MAPLOOP}, match={XXX}, replace={ETZ Arch.}]
			\step[fieldsource=\regexp{$MAPLOOP}, match={XXX}, replace={Feingeraetetechnik}]
			\step[fieldsource=\regexp{$MAPLOOP}, match={XXX}, replace={Feinw. Tech. Messtech.}]
			\step[fieldsource=\regexp{$MAPLOOP}, match={XXX}, replace={Fert. Tech. Betr.}]
			\step[fieldsource=\regexp{$MAPLOOP}, match={XXX}, replace={Fis. Tecnol.}]
			\step[fieldsource=\regexp{$MAPLOOP}, match={XXX}, replace={Fiz. Khim. Obrab. Mater.}]
			\step[fieldsource=\regexp{$MAPLOOP}, match={XXX}, replace={Fiz. Met. Metalloved}]
			\step[fieldsource=\regexp{$MAPLOOP}, match={XXX}, replace={Fiz. Nizk. Temp.}]
			\step[fieldsource=\regexp{$MAPLOOP}, match={XXX}, replace={Fiz. Plazmy}]
			\step[fieldsource=\regexp{$MAPLOOP}, match={XXX}, replace={Fiz. Tekh. Poluprovodn.}]
			\step[fieldsource=\regexp{$MAPLOOP}, match={XXX}, replace={Fiz. Tverd. Tela}]
			\step[fieldsource=\regexp{$MAPLOOP}, match={XXX}, replace={Fiz.-Khim. Mekh. Mater.}]
			\step[fieldsource=\regexp{$MAPLOOP}, match={XXX}, replace={Fizika}]
			\step[fieldsource=\regexp{$MAPLOOP}, match={XXX}, replace={Forsch.-Ber. Landes Nordrh.-Westfal.}]
			\step[fieldsource=\regexp{$MAPLOOP}, match={XXX}, replace={Frequenz}]
			\step[fieldsource=\regexp{$MAPLOOP}, match={XXX}, replace={Fys. Tidsskr.}]
			\step[fieldsource=\regexp{$MAPLOOP}, match={XXX}, replace={G. Fis.}]
			\step[fieldsource=\regexp{$MAPLOOP}, match={XXX}, replace={Geliotekhnika}]
			\step[fieldsource=\regexp{$MAPLOOP}, match={XXX}, replace={Geochim. Cosmochim. Acta}]
			\step[fieldsource=\regexp{$MAPLOOP}, match={XXX}, replace={Haerterei-Tech. Mitt.}]
			\step[fieldsource=\regexp{$MAPLOOP}, match={XXX}, replace={Helv. Chim. Acta}]
			\step[fieldsource=\regexp{$MAPLOOP}, match={XXX}, replace={Helv. Med. Acta}]
			\step[fieldsource=\regexp{$MAPLOOP}, match={XXX}, replace={Helv. Phys. Acta}]
			\step[fieldsource=\regexp{$MAPLOOP}, match={XXX}, replace={Hochfreq. Electroakust}]
			\step[fieldsource=\regexp{$MAPLOOP}, match={XXX}, replace={Hoppe-Seylers Z. Physiol. Chem.}]
			\step[fieldsource=\regexp{$MAPLOOP}, match={XXX}, replace={Inf. Elektron.}]
			\step[fieldsource=\regexp{$MAPLOOP}, match={XXX}, replace={Inf. Elettron.}]
			\step[fieldsource=\regexp{$MAPLOOP}, match={XXX}, replace={Inform. Forsch. Entwickl.}]
			\step[fieldsource=\regexp{$MAPLOOP}, match={XXX}, replace={Inform. Spektrum}]
			\step[fieldsource=\regexp{$MAPLOOP}, match={XXX}, replace={Inform.-Fachber.}]
			\step[fieldsource=\regexp{$MAPLOOP}, match={XXX}, replace={Informatie}]
			\step[fieldsource=\regexp{$MAPLOOP}, match={XXX}, replace={Informatik}]
			\step[fieldsource=\regexp{$MAPLOOP}, match={XXX}, replace={Informatologia Yugosl.}]
			\step[fieldsource=\regexp{$MAPLOOP}, match={XXX}, replace={Informatyka}]
			\step[fieldsource=\regexp{$MAPLOOP}, match={XXX}, replace={Infowelt}]
			\step[fieldsource=\regexp{$MAPLOOP}, match={XXX}, replace={Ing. Electr. Mec.}]
			\step[fieldsource=\regexp{$MAPLOOP}, match={XXX}, replace={Ing. Mec. Electr.}]
			\step[fieldsource=\regexp{$MAPLOOP}, match={XXX}, replace={Ing.-Arch.}]
			\step[fieldsource=\regexp{$MAPLOOP}, match={XXX}, replace={Inzh.-Fiz. Zh.}]
			\step[fieldsource=\regexp{$MAPLOOP}, match={XXX}, replace={Izmer. Tekh.}]
			\step[fieldsource=\regexp{$MAPLOOP}, match={XXX}, replace={Izv. Akad. Nauk Arm. SSR Ser. Tekh. Nauk}]
			\step[fieldsource=\regexp{$MAPLOOP}, match={XXX}, replace={Izv. Akad. Nauk SSSR Energ. Transp.}]
			\step[fieldsource=\regexp{$MAPLOOP}, match={XXX}, replace={Izv. Akad. Nauk SSSR Fiz. Atmos. Okeana}]
			\step[fieldsource=\regexp{$MAPLOOP}, match={XXX}, replace={Izv. Akad. Nauk SSSR Fiz. Zemli}]
			\step[fieldsource=\regexp{$MAPLOOP}, match={XXX}, replace={Izv. Akad. Nauk SSSR Ser. Fiz.}]
			\step[fieldsource=\regexp{$MAPLOOP}, match={XXX}, replace={Izv. Vyssh. Uchebn. Zaved. Elektromekh.}]
			\step[fieldsource=\regexp{$MAPLOOP}, match={XXX}, replace={Izv. Vyssh. Uchebn. Zaved. Radioelektron.}]
			\step[fieldsource=\regexp{$MAPLOOP}, match={XXX}, replace={Izv. Vyssh. Uchebn. Zaved. Radiofiz.}]
			\step[fieldsource=\regexp{$MAPLOOP}, match={XXX}, replace={J. Chim Phys. Phys.-Chim Biol.}]
			\step[fieldsource=\regexp{$MAPLOOP}, match={XXX}, replace={Kernenergie}]
			\step[fieldsource=\regexp{$MAPLOOP}, match={XXX}, replace={Kerntechnik}]
			\step[fieldsource=\regexp{$MAPLOOP}, match={XXX}, replace={Khim. Fiz.}]
			\step[fieldsource=\regexp{$MAPLOOP}, match={XXX}, replace={Kibern. Vychisl. Tekh.}]
			\step[fieldsource=\regexp{$MAPLOOP}, match={XXX}, replace={Kibernetika}]
			\step[fieldsource=\regexp{$MAPLOOP}, match={XXX}, replace={Kristallografiya}]
			\step[fieldsource=\regexp{$MAPLOOP}, match={XXX}, replace={Kvantovaya Elektron. Mosk.}]
			\step[fieldsource=\regexp{$MAPLOOP}, match={XXX}, replace={Kybernetes}]
			\step[fieldsource=\regexp{$MAPLOOP}, match={XXX}, replace={Kybernetika}]
			\step[fieldsource=\regexp{$MAPLOOP}, match={XXX}, replace={Med. Tek.}]
			\step[fieldsource=\regexp{$MAPLOOP}, match={XXX}, replace={Mekh. Avtom. Proizvod.}]
			\step[fieldsource=\regexp{$MAPLOOP}, match={XXX}, replace={Meres Autom.}]
			\step[fieldsource=\regexp{$MAPLOOP}, match={XXX}, replace={Mesures}]
			\step[fieldsource=\regexp{$MAPLOOP}, match={XXX}, replace={Metallofizika}]
			\step[fieldsource=\regexp{$MAPLOOP}, match={XXX}, replace={Metalloved. Term. Obrab. Met.}]
			\step[fieldsource=\regexp{$MAPLOOP}, match={XXX}, replace={Meteorol. Gidrol.}]
			\step[fieldsource=\regexp{$MAPLOOP}, match={XXX}, replace={Meteorol. Rundsch.}]
			\step[fieldsource=\regexp{$MAPLOOP}, match={XXX}, replace={Metrol. Apl.}]
			\step[fieldsource=\regexp{$MAPLOOP}, match={XXX}, replace={Metrologia}]
			\step[fieldsource=\regexp{$MAPLOOP}, match={XXX}, replace={Medel. Simul.}]
			\step[fieldsource=\regexp{$MAPLOOP}, match={XXX}, replace={Nachr. Dok.}]
			\step[fieldsource=\regexp{$MAPLOOP}, match={XXX}, replace={Nachr.tech. Elektron.}]
			\step[fieldsource=\regexp{$MAPLOOP}, match={XXX}, replace={Naturwissentchaften}]
			\step[fieldsource=\regexp{$MAPLOOP}, match={XXX}, replace={Neue Tech.}]
			\step[fieldsource=\regexp{$MAPLOOP}, match={XXX}, replace={Neue Tech. Buero}]
			\step[fieldsource=\regexp{$MAPLOOP}, match={XXX}, replace={Nukleonika}]
			\step[fieldsource=\regexp{$MAPLOOP}, match={XXX}, replace={Numer. Math.}]
			\step[fieldsource=\regexp{$MAPLOOP}, match={XXX}, replace={Nuovo Cimento A}]
			\step[fieldsource=\regexp{$MAPLOOP}, match={XXX}, replace={Nuovo Cimento B}]
			\step[fieldsource=\regexp{$MAPLOOP}, match={XXX}, replace={Nouvo Cimento C}]
			\step[fieldsource=\regexp{$MAPLOOP}, match={XXX}, replace={Nuovo Cimento D}]
			\step[fieldsource=\regexp{$MAPLOOP}, match={XXX}, replace={Okeanologiya}]
			\step[fieldsource=\regexp{$MAPLOOP}, match={XXX}, replace={Opt. Spektrosk.}]
			\step[fieldsource=\regexp{$MAPLOOP}, match={XXX}, replace={Opt.-Mekh. Prom.}]
			\step[fieldsource=\regexp{$MAPLOOP}, match={XXX}, replace={Optik}]
			\step[fieldsource=\regexp{$MAPLOOP}, match={XXX}, replace={Photogrammetria}]
			\step[fieldsource=\regexp{$MAPLOOP}, match={XXX}, replace={Photonics Spectra}]
			\step[fieldsource=\regexp{$MAPLOOP}, match={XXX}, replace={Pis’ma Astron. Zh. }]
			\step[fieldsource=\regexp{$MAPLOOP}, match={XXX}, replace={Pis’ma Zh. Eksp. Teor. Fiz. }]
			\step[fieldsource=\regexp{$MAPLOOP}, match={XXX}, replace={Pis’ma Zh. Tekh. Fiz. }]
			\step[fieldsource=\regexp{$MAPLOOP}, match={XXX}, replace={Poverkhn., Fiz. Khim. Mekh.}]
			\step[fieldsource=\regexp{$MAPLOOP}, match={XXX}, replace={Pr. Inst. Elektrotech.}]
			\step[fieldsource=\regexp{$MAPLOOP}, match={XXX}, replace={Prib. Sist. Upr.}]
			\step[fieldsource=\regexp{$MAPLOOP}, match={XXX}, replace={Prib. Tekh. Eksp.}]
			\step[fieldsource=\regexp{$MAPLOOP}, match={XXX}, replace={Prikl. Mat. Mekh.}]
			\step[fieldsource=\regexp{$MAPLOOP}, match={XXX}, replace={Prikl. Mekh.}]
			\step[fieldsource=\regexp{$MAPLOOP}, match={XXX}, replace={Probl. Kibern.}]
			\step[fieldsource=\regexp{$MAPLOOP}, match={XXX}, replace={Probl. Peredachi Inf.}]
			\step[fieldsource=\regexp{$MAPLOOP}, match={XXX}, replace={Proc. K. Ned. Akad. Wet. B, Palaeontol.}]
			\step[fieldsource=\regexp{$MAPLOOP}, match={XXX}, replace={Anthropol. }]
			\step[fieldsource=\regexp{$MAPLOOP}, match={XXX}, replace={Programmirovanie}]
			\step[fieldsource=\regexp{$MAPLOOP}, match={XXX}, replace={Prz. Elektrotech.}]
			\step[fieldsource=\regexp{$MAPLOOP}, match={XXX}, replace={Prz. Telekomun.}]
			\step[fieldsource=\regexp{$MAPLOOP}, match={XXX}, replace={PT/Elektrotech. Elektron.}]
			\step[fieldsource=\regexp{$MAPLOOP}, match={XXX}, replace={Radio Fernsehen Elektron.}]
			\step[fieldsource=\regexp{$MAPLOOP}, match={XXX}, replace={Radiotekh. Elektron.}]
			\step[fieldsource=\regexp{$MAPLOOP}, match={XXX}, replace={Radiotekhnika Mosk.}]
			\step[fieldsource=\regexp{$MAPLOOP}, match={XXX}, replace={Rev. Acad. Cienc. Zaragoza}]
			\step[fieldsource=\regexp{$MAPLOOP}, match={XXX}, replace={Rev. Electrotec. (Argentina)}]
			\step[fieldsource=\regexp{$MAPLOOP}, match={XXX}, replace={Rev. Electrotec. (Spain)}]
			\step[fieldsource=\regexp{$MAPLOOP}, match={XXX}, replace={Rev. Energ.}]
			\step[fieldsource=\regexp{$MAPLOOP}, match={XXX}, replace={Rev. Esp. Electron. Rev. Geofis.}]
			\step[fieldsource=\regexp{$MAPLOOP}, match={XXX}, replace={Ric. Autom.}]
			\step[fieldsource=\regexp{$MAPLOOP}, match={XXX}, replace={Ric. Spettrosc.}]
			\step[fieldsource=\regexp{$MAPLOOP}, match={XXX}, replace={Robotersysteme}]
			\step[fieldsource=\regexp{$MAPLOOP}, match={XXX}, replace={Rozpr. Electrotech.}]
			\step[fieldsource=\regexp{$MAPLOOP}, match={XXX}, replace={Sadhana}]
			\step[fieldsource=\regexp{$MAPLOOP}, match={XXX}, replace={Schweiz. Tech. Z.}]
			\step[fieldsource=\regexp{$MAPLOOP}, match={XXX}, replace={Scientia}]
			\step[fieldsource=\regexp{$MAPLOOP}, match={XXX}, replace={Siemens Forsch. Entwickl. Ber.}]
			\step[fieldsource=\regexp{$MAPLOOP}, match={XXX}, replace={Sist. Autom.}]
			\step[fieldsource=\regexp{$MAPLOOP}, match={XXX}, replace={Sitzungsber. Oester. Akad. Wiss. Math.-}]
			\step[fieldsource=\regexp{$MAPLOOP}, match={XXX}, replace={Naturwiss. Kl. Abt. II (Austria)}]
			\step[fieldsource=\regexp{$MAPLOOP}, match={XXX}, replace={Spectrochim. Acta A, Mol. Spectrosc.}]
			\step[fieldsource=\regexp{$MAPLOOP}, match={XXX}, replace={Spectrochim. Acta B, At. Spectrosc.}]
			\step[fieldsource=\regexp{$MAPLOOP}, match={XXX}, replace={Sprache Datenverarb.}]
			\step[fieldsource=\regexp{$MAPLOOP}, match={XXX}, replace={Stanki Instrum.}]
			\step[fieldsource=\regexp{$MAPLOOP}, match={XXX}, replace={Steklo Keram.}]
			\step[fieldsource=\regexp{$MAPLOOP}, match={XXX}, replace={Svetotekhnika  TE Int.}]
			\step[fieldsource=\regexp{$MAPLOOP}, match={XXX}, replace={Tech. Bull. Vevey}]
			\step[fieldsource=\regexp{$MAPLOOP}, match={XXX}, replace={Tech. Mitt. Krupp (Engl. Ed.)}]
			\step[fieldsource=\regexp{$MAPLOOP}, match={XXX}, replace={Tech. Mitt. PTT}]
			\step[fieldsource=\regexp{$MAPLOOP}, match={XXX}, replace={Tech. Mitt. RFZ}]
			\step[fieldsource=\regexp{$MAPLOOP}, match={XXX}, replace={Technica}]
			\step[fieldsource=\regexp{$MAPLOOP}, match={XXX}, replace={Tecnica}]
			\step[fieldsource=\regexp{$MAPLOOP}, match={XXX}, replace={Teh. Fiz.}]
			\step[fieldsource=\regexp{$MAPLOOP}, match={XXX}, replace={Tehnika}]
			\step[fieldsource=\regexp{$MAPLOOP}, match={XXX}, replace={Tekh. Elektrodin.}]
			\step[fieldsource=\regexp{$MAPLOOP}, match={XXX}, replace={Tekh. Kibern.}]
			\step[fieldsource=\regexp{$MAPLOOP}, match={XXX}, replace={Tekh. Kino Telev.}]
			\step[fieldsource=\regexp{$MAPLOOP}, match={XXX}, replace={Tekh. Misul}]
			\step[fieldsource=\regexp{$MAPLOOP}, match={XXX}, replace={Telekomunikacije}]
			\step[fieldsource=\regexp{$MAPLOOP}, match={XXX}, replace={Telektronikk}]
			\step[fieldsource=\regexp{$MAPLOOP}, match={XXX}, replace={Teleteknik}]
			\step[fieldsource=\regexp{$MAPLOOP}, match={XXX}, replace={Teor. Mat. Fiz.}]
			\step[fieldsource=\regexp{$MAPLOOP}, match={XXX}, replace={Teploeoergetika}]
			\step[fieldsource=\regexp{$MAPLOOP}, match={XXX}, replace={Teplofiz. Vvs. Temp.}]
			\step[fieldsource=\regexp{$MAPLOOP}, match={XXX}, replace={Tidskr. Dok.}]
			\step[fieldsource=\regexp{$MAPLOOP}, match={XXX}, replace={TN Nachr.}]
			\step[fieldsource=\regexp{$MAPLOOP}, match={XXX}, replace={Toute Electron.}]
			\step[fieldsource=\regexp{$MAPLOOP}, match={XXX}, replace={Tr. Inst. Teor. Astron.}]
			\step[fieldsource=\regexp{$MAPLOOP}, match={XXX}, replace={Ukr. Fiz. Zh.}]
			\step[fieldsource=\regexp{$MAPLOOP}, match={XXX}, replace={Usp. Fiz. Nauk}]
			\step[fieldsource=\regexp{$MAPLOOP}, match={XXX}, replace={Vak.-Tech.}]
			\step[fieldsource=\regexp{$MAPLOOP}, match={XXX}, replace={VDE Fachiber.}]
			\step[fieldsource=\regexp{$MAPLOOP}, match={XXX}, replace={VDI Z.}]
			\step[fieldsource=\regexp{$MAPLOOP}, match={XXX}, replace={Vestn. Mashinostr.}]
			\step[fieldsource=\regexp{$MAPLOOP}, match={XXX}, replace={Vestn. Mosk. Univ. 15, Vychisl. Mat. Kibern.}]
			\step[fieldsource=\regexp{$MAPLOOP}, match={XXX}, replace={Vestn. Mosk. Univ. 3, Fiz. Astron.}]
			\step[fieldsource=\regexp{$MAPLOOP}, match={XXX}, replace={Vesti Akad. Navuk BSSR Ser. Fiz. Energ. Navuk}]
			\step[fieldsource=\regexp{$MAPLOOP}, match={XXX}, replace={VGB Kraftwerkstech. (Ger. Ed.)}]
			\step[fieldsource=\regexp{$MAPLOOP}, match={XXX}, replace={Vide Couches Minces}]
			\step[fieldsource=\regexp{$MAPLOOP}, match={XXX}, replace={Vistas Astron.}]
			\step[fieldsource=\regexp{$MAPLOOP}, match={XXX}, replace={Vopr. At. Nauki Tekh. Ser., Fiz. Radiats.}]
			\step[fieldsource=\regexp{$MAPLOOP}, match={XXX}, replace={Povrezhdenii Radiats. Materialoved.}]
			\step[fieldsource=\regexp{$MAPLOOP}, match={XXX}, replace={Vopr. At. Nauki Tekh. Ser., Obshch. Yad. Fiz.}]
			\step[fieldsource=\regexp{$MAPLOOP}, match={XXX}, replace={Vuoto Sci. Tecnol.}]
			\step[fieldsource=\regexp{$MAPLOOP}, match={XXX}, replace={Wiss. Z. Friedrich-Schiller-Univ. Jena}]
			\step[fieldsource=\regexp{$MAPLOOP}, match={XXX}, replace={Nat.wiss. Reihe}]
			\step[fieldsource=\regexp{$MAPLOOP}, match={XXX}, replace={Wiss. Z. Karl-Marx-Univ. Leipz. Math.-}]
			\step[fieldsource=\regexp{$MAPLOOP}, match={XXX}, replace={Nat.wiss. Reihe}]
			\step[fieldsource=\regexp{$MAPLOOP}, match={XXX}, replace={Wiss. Z. Tech. Hochsch. Ilmenau}]
			\step[fieldsource=\regexp{$MAPLOOP}, match={XXX}, replace={Wiss. Z. Tech. Univ. Dresd.}]
			\step[fieldsource=\regexp{$MAPLOOP}, match={XXX}, replace={Wiss. Z. Tech. Univ. Karl-Marx-Stadt}]
			\step[fieldsource=\regexp{$MAPLOOP}, match={XXX}, replace={Yad. Fiz.}]
			\step[fieldsource=\regexp{$MAPLOOP}, match={XXX}, replace={Z. Angew. Math. Mech.}]
			\step[fieldsource=\regexp{$MAPLOOP}, match={XXX}, replace={Z. Angew. Math. Phys.}]
			\step[fieldsource=\regexp{$MAPLOOP}, match={XXX}, replace={Z. Met.kd.}]
			\step[fieldsource=\regexp{$MAPLOOP}, match={XXX}, replace={Z. Nat. Forsch. A, Phys. Phys. Chem. Kosmophys.}]
			\step[fieldsource=\regexp{$MAPLOOP}, match={XXX}, replace={Z. Oper. Res. A, Theor.}]
			\step[fieldsource=\regexp{$MAPLOOP}, match={XXX}, replace={Z. Oper. Res. B, Prax.}]
			\step[fieldsource=\regexp{$MAPLOOP}, match={XXX}, replace={Z. Phys.}]
			\step[fieldsource=\regexp{$MAPLOOP}, match={XXX}, replace={Z. Phys. A, At. Nuclei}]
			\step[fieldsource=\regexp{$MAPLOOP}, match={XXX}, replace={Z. Phys. B, Condens. Matter}]
			\step[fieldsource=\regexp{$MAPLOOP}, match={XXX}, replace={Z. Phys. C, Part Fields}]
			\step[fieldsource=\regexp{$MAPLOOP}, match={XXX}, replace={Z. Phys. Chem. Neue Folge}]
			\step[fieldsource=\regexp{$MAPLOOP}, match={XXX}, replace={Z. Phys. Chem., Leipz.}]
			\step[fieldsource=\regexp{$MAPLOOP}, match={XXX}, replace={Z. Phys. D, At. Mol. Clusters}]
			\step[fieldsource=\regexp{$MAPLOOP}, match={XXX}, replace={Zavod. Lab.}]
			\step[fieldsource=\regexp{$MAPLOOP}, match={XXX}, replace={Zh. Eksp. Teor. Fiz.}]
			\step[fieldsource=\regexp{$MAPLOOP}, match={XXX}, replace={Zh. Fiz. Khim.}]
			\step[fieldsource=\regexp{$MAPLOOP}, match={XXX}, replace={Zh. Prikl. Mekh. Tekh. Fiz.}]
			\step[fieldsource=\regexp{$MAPLOOP}, match={XXX}, replace={Zh. Prikl. Spektrosk.}]
			\step[fieldsource=\regexp{$MAPLOOP}, match={XXX}, replace={Zh. Tekh. Fiz.}]
			\step[fieldsource=\regexp{$MAPLOOP}, match={XXX}, replace={Zh. Vychisl. Mat. Mat. Fiz.}]
			\step[fieldsource=\regexp{$MAPLOOP}, match={XXX}, replace={Zisin, J. Seismol. Soc. Jpn.}] 
			\step[fieldsource=\regexp{$MAPLOOP}, match={Production}, replace={Prod.}]
			\step[fieldsource=\regexp{$MAPLOOP}, match={Reliability}, replace={Rel.}]
			\step[fieldsource=\regexp{$MAPLOOP}, match={Report}, replace={Rep.}]
			\step[fieldsource=\regexp{$MAPLOOP}, match={Semiconductor}, replace={Semicond.}]
			\step[fieldsource=\regexp{$MAPLOOP}, match={Research}, replace={Res.}]
			\step[fieldsource=\regexp{$MAPLOOP}, match={Sensing}, replace={Sens.}]
			\step[fieldsource=\regexp{$MAPLOOP}, match={Resonance}, replace={Reson.}]
			\step[fieldsource=\regexp{$MAPLOOP}, match={Series}, replace={Ser.}]
			\step[fieldsource=\regexp{$MAPLOOP}, match={Resources}, replace={Resour.}]
			\step[fieldsource=\regexp{$MAPLOOP}, match={Simulation}, replace={Simul.}]
			\step[fieldsource=\regexp{$MAPLOOP}, match={Reviews}, replace={Rev.}]
			\step[fieldsource=\regexp{$MAPLOOP}, match={Review}, replace={Rev.}]
			\step[fieldsource=\regexp{$MAPLOOP}, match={Singapore}, replace={Singap.}]
			\step[fieldsource=\regexp{$MAPLOOP}, match={Robotics}, replace={Robot.}]
			\step[fieldsource=\regexp{$MAPLOOP}, match={Sistema}, replace={Sist.}]
			\step[fieldsource=\regexp{$MAPLOOP}, match={Royal}, replace={Roy.}]
			\step[fieldsource=\regexp{$MAPLOOP}, match={Society}, replace={Soc.}]
			\step[fieldsource=\regexp{$MAPLOOP}, match={Safety}, replace={Saf.}]
			\step[fieldsource=\regexp{$MAPLOOP}, match={Sociological}, replace={Sociol.}]
			\step[fieldsource=\regexp{$MAPLOOP}, match={Satellite}, replace={Satell.}]
			\step[fieldsource=\regexp{$MAPLOOP}, match={Software}, replace={Softw.}]
			\step[fieldsource=\regexp{$MAPLOOP}, match={Scandinavian}, replace={Scand.}]
			\step[fieldsource=\regexp{$MAPLOOP}, match={Solar}, replace={Sol.}]
			\step[fieldsource=\regexp{$MAPLOOP}, match={Sciences}, replace={Sci.}]
			\step[fieldsource=\regexp{$MAPLOOP}, match={Science}, replace={Sci.}]
			\step[fieldsource=\regexp{$MAPLOOP}, match={Soviet}, replace={Sov.}]
			\step[fieldsource=\regexp{$MAPLOOP}, match={Section}, replace={Sect.}]
			\step[fieldsource=\regexp{$MAPLOOP}, match={Spectroscopy}, replace={Spectrosc.}]
			\step[fieldsource=\regexp{$MAPLOOP}, match={Security}, replace={Secur.}]
			\step[fieldsource=\regexp{$MAPLOOP}, match={Spectrum}, replace={Spectr.}]
			\step[fieldsource=\regexp{$MAPLOOP}, match={Seismology}, replace={Seismol.}]
			\step[fieldsource=\regexp{$MAPLOOP}, match={Speculations}, replace={Specul.}]
			\step[fieldsource=\regexp{$MAPLOOP}, match={Selected}, replace={Sel.}]
			\step[fieldsource=\regexp{$MAPLOOP}, match={Statistics}, replace={Statist.}]
			\step[fieldsource=\regexp{$MAPLOOP}, match={Structures}, replace={Struct.}]
			\step[fieldsource=\regexp{$MAPLOOP}, match={Structure}, replace={Struct.}]
			\step[fieldsource=\regexp{$MAPLOOP}, match={Terrestrial}, replace={Terr.}]
			\step[fieldsource=\regexp{$MAPLOOP}, match={Studies}, replace={Stud.}]
			\step[fieldsource=\regexp{$MAPLOOP}, match={Theoretical}, replace={Theor.}]
			\step[fieldsource=\regexp{$MAPLOOP}, match={Superconductivity}, replace={Supercond.}]
			\step[fieldsource=\regexp{$MAPLOOP}, match={Transactions}, replace={Trans.}]
			\step[fieldsource=\regexp{$MAPLOOP}, match={Supplement}, replace={Suppl.}]
			\step[fieldsource=\regexp{$MAPLOOP}, match={Translation}, replace={Transl.}]
			\step[fieldsource=\regexp{$MAPLOOP}, match={Surface}, replace={Surf.}]
			\step[fieldsource=\regexp{$MAPLOOP}, match={Transmission}, replace={Transmiss.}]
			\step[fieldsource=\regexp{$MAPLOOP}, match={Survey}, replace={Surv.}]
			\step[fieldsource=\regexp{$MAPLOOP}, match={Transportation}, replace={Transp.}]
			\step[fieldsource=\regexp{$MAPLOOP}, match={Sustainable}, replace={Sustain.}]
			\step[fieldsource=\regexp{$MAPLOOP}, match={Tutorials}, replace={Tut.}]
			\step[fieldsource=\regexp{$MAPLOOP}, match={Symposium}, replace={Symp.}]
			\step[fieldsource=\regexp{$MAPLOOP}, match={Ultrasonic}, replace={Ultrason.}]
			\step[fieldsource=\regexp{$MAPLOOP}, match={Systems}, replace={Syst.}]
			\step[fieldsource=\regexp{$MAPLOOP}, match={System}, replace={Syst.}]
			\step[fieldsource=\regexp{$MAPLOOP}, match={University}, replace={Univ.}]
			\step[fieldsource=\regexp{$MAPLOOP}, match={\detokenize{Universität}}, replace={Univ.}]
			\step[fieldsource=\regexp{$MAPLOOP}, match={\detokenize{Université}}, replace={Univ.}]
			\step[fieldsource=\regexp{$MAPLOOP}, match={Vacuum}, replace={Vac.}]
			\step[fieldsource=\regexp{$MAPLOOP}, match={Vehicles}, replace={Veh.}]
			\step[fieldsource=\regexp{$MAPLOOP}, match={Vehicle}, replace={Veh.}]
			\step[fieldsource=\regexp{$MAPLOOP}, match={Vehicular}, replace={Veh.}]
			\step[fieldsource=\regexp{$MAPLOOP}, match={Technology}, replace={Technol.}]
			\step[fieldsource=\regexp{$MAPLOOP}, match={Technological}, replace={Technol.}]
			\step[fieldsource=\regexp{$MAPLOOP}, match={Vibration}, replace={Vib.}]
			\step[fieldsource=\regexp{$MAPLOOP}, match={Telecommunications}, replace={Telecommun.}]
			\step[fieldsource=\regexp{$MAPLOOP}, match={Visual}, replace={Vis.}]
			\step[fieldsource=\regexp{$MAPLOOP}, match={Television}, replace={Telev.}]
			\step[fieldsource=\regexp{$MAPLOOP}, match={Welding}, replace={Weld.}]
			\step[fieldsource=\regexp{$MAPLOOP}, match={Temperature}, replace={Temp.}]
			\step[fieldsource=\regexp{$MAPLOOP}, match={Working}, replace={Work.}]
			\step[fieldsource=\regexp{$MAPLOOP}, match={Learning}, replace={Learn.}]
			\step[fieldsource=\regexp{$MAPLOOP}, match={Measurement}, replace={Meas.}]
			\step[fieldsource=\regexp{$MAPLOOP}, match={Letters}, replace={Lett.}]
			\step[fieldsource=\regexp{$MAPLOOP}, match={Letter}, replace={Lett.}]
			\step[fieldsource=\regexp{$MAPLOOP}, match={Mechanical}, replace={Mech.}]
			\step[fieldsource=\regexp{$MAPLOOP}, match={Mechanics}, replace={Mech.}]
			\step[fieldsource=\regexp{$MAPLOOP}, match={Mechanic}, replace={Mech.}]
			\step[fieldsource=\regexp{$MAPLOOP}, match={Lightwave}, replace={Lightw.}]
			\step[fieldsource=\regexp{$MAPLOOP}, match={Medical}, replace={Med.}]
			\step[fieldsource=\regexp{$MAPLOOP}, match={Logic, Logical}, replace={Log.}]
			\step[fieldsource=\regexp{$MAPLOOP}, match={Metals}, replace={Met.}]
			\step[fieldsource=\regexp{$MAPLOOP}, match={Luminescence}, replace={Lumin.}]
			\step[fieldsource=\regexp{$MAPLOOP}, match={Metallurgy}, replace={Metall.}]
			\step[fieldsource=\regexp{$MAPLOOP}, match={Machines}, replace={Mach.}]
			\step[fieldsource=\regexp{$MAPLOOP}, match={Machine}, replace={Mach.}]
			\step[fieldsource=\regexp{$MAPLOOP}, match={Meteorology}, replace={Meteorol.}]
			\step[fieldsource=\regexp{$MAPLOOP}, match={Magazine}, replace={Mag.}]
			\step[fieldsource=\regexp{$MAPLOOP}, match={Metropolitan}, replace={Metrop.}]
			\step[fieldsource=\regexp{$MAPLOOP}, match={Magnetics}, replace={Magn.}]
			\step[fieldsource=\regexp{$MAPLOOP}, match={Mexican, Mexico}, replace={Mex.}]
			\step[fieldsource=\regexp{$MAPLOOP}, match={Management}, replace={Manage.}]
			\step[fieldsource=\regexp{$MAPLOOP}, match={Microelectromechanical}, replace={Microelectromech.}]
			\step[fieldsource=\regexp{$MAPLOOP}, match={Managing}, replace={Manag.}]
			\step[fieldsource=\regexp{$MAPLOOP}, match={Microgravity}, replace={Microgr.}]
			\step[fieldsource=\regexp{$MAPLOOP}, match={Manufacturing}, replace={Manuf.}]
			\step[fieldsource=\regexp{$MAPLOOP}, match={Microscopy}, replace={Microsc.}]
			\step[fieldsource=\regexp{$MAPLOOP}, match={Marine}, replace={Mar.}]
			\step[fieldsource=\regexp{$MAPLOOP}, match={Microwaves}, replace={Microw.}]
			\step[fieldsource=\regexp{$MAPLOOP}, match={Microwave}, replace={Microw.}]
			\step[fieldsource=\regexp{$MAPLOOP}, match={Materials}, replace={Mater.}]
			\step[fieldsource=\regexp{$MAPLOOP}, match={Material}, replace={Mater.}]
			\step[fieldsource=\regexp{$MAPLOOP}, match={Military}, replace={Mil.}]
			\step[fieldsource=\regexp{$MAPLOOP}, match={Modeling}, replace={Model.}]
			\step[fieldsource=\regexp{$MAPLOOP}, match={Modelling}, replace={Model.}]
			\step[fieldsource=\regexp{$MAPLOOP}, match={Oceanic}, replace={Ocean.}]
			\step[fieldsource=\regexp{$MAPLOOP}, match={Molecular}, replace={Mol.}]
			\step[fieldsource=\regexp{$MAPLOOP}, match={Oceanography}, replace={Oceanogr.}]
			\step[fieldsource=\regexp{$MAPLOOP}, match={Monitoring}, replace={Monit.}]
			\step[fieldsource=\regexp{$MAPLOOP}, match={Occupation}, replace={Occupat.}]
			\step[fieldsource=\regexp{$MAPLOOP}, match={Multiphysics}, replace={Multiphys.}]
			\step[fieldsource=\regexp{$MAPLOOP}, match={Operational}, replace={Oper.}]
			\step[fieldsource=\regexp{$MAPLOOP}, match={Nanobioscience}, replace={Nanobiosci.}]
			\step[fieldsource=\regexp{$MAPLOOP}, match={Optical}, replace={Opt.}]
			\step[fieldsource=\regexp{$MAPLOOP}, match={Nanotechnology}, replace={Nanotechnol.}]
			\step[fieldsource=\regexp{$MAPLOOP}, match={Optics}, replace={Opt.}]
			\step[fieldsource=\regexp{$MAPLOOP}, match={National}, replace={Nat.}]
			\step[fieldsource=\regexp{$MAPLOOP}, match={Optimization}, replace={Optim.}]
			\step[fieldsource=\regexp{$MAPLOOP}, match={Naval}, replace={Nav.}]
			\step[fieldsource=\regexp{$MAPLOOP}, match={Organization}, replace={Org.}]
			\step[fieldsource=\regexp{$MAPLOOP}, match={Networking}, replace={Netw.}]
			\step[fieldsource=\regexp{$MAPLOOP}, match={Networked}, replace={Netw.}]
			\step[fieldsource=\regexp{$MAPLOOP}, match={Network}, replace={Netw.}]
			\step[fieldsource=\regexp{$MAPLOOP}, match={Packaging}, replace={Packag.}]
			\step[fieldsource=\regexp{$MAPLOOP}, match={Newsletter}, replace={Newslett.}]
			\step[fieldsource=\regexp{$MAPLOOP}, match={Particle}, replace={Part.}]
			\step[fieldsource=\regexp{$MAPLOOP}, match={Nondestructive}, replace={Nondestruct.}]
			\step[fieldsource=\regexp{$MAPLOOP}, match={Patent}, replace={Pat.}]
			\step[fieldsource=\regexp{$MAPLOOP}, match={Nuclear}, replace={Nucl.}]
			\step[fieldsource=\regexp{$MAPLOOP}, match={Performance}, replace={Perform.}]
			\step[fieldsource=\regexp{$MAPLOOP}, match={Numerical}, replace={Numer.}]
			\step[fieldsource=\regexp{$MAPLOOP}, match={Personal}, replace={Pers.}]
			\step[fieldsource=\regexp{$MAPLOOP}, match={Observations}, replace={Observ.}]
			\step[fieldsource=\regexp{$MAPLOOP}, match={Philosophical}, replace={Philos.}]
			\step[fieldsource=\regexp{$MAPLOOP}, match={Photonics}, replace={Photon.}]
			\step[fieldsource=\regexp{$MAPLOOP}, match={Productivity}, replace={Productiv.}]
			\step[fieldsource=\regexp{$MAPLOOP}, match={Photovoltaics}, replace={Photovolt.}]
			\step[fieldsource=\regexp{$MAPLOOP}, match={Programming}, replace={Program.}]
			\step[fieldsource=\regexp{$MAPLOOP}, match={Physics}, replace={Phys.}]
			\step[fieldsource=\regexp{$MAPLOOP}, match={Progress}, replace={Prog.}]
			\step[fieldsource=\regexp{$MAPLOOP}, match={Physiology}, replace={Physiol.}]
			\step[fieldsource=\regexp{$MAPLOOP}, match={Propagation}, replace={Propag.}]
			\step[fieldsource=\regexp{$MAPLOOP}, match={Planetary}, replace={Planet.}]
			\step[fieldsource=\regexp{$MAPLOOP}, match={Psychology}, replace={Psychol.}]
			\step[fieldsource=\regexp{$MAPLOOP}, match={Pneumatics}, replace={Pneum.}]
			\step[fieldsource=\regexp{$MAPLOOP}, match={Quality}, replace={Qual.}]
			\step[fieldsource=\regexp{$MAPLOOP}, match={Pollution}, replace={Pollut.}]
			\step[fieldsource=\regexp{$MAPLOOP}, match={Quarterly}, replace={Quart.}]
			\step[fieldsource=\regexp{$MAPLOOP}, match={Polymer}, replace={Polym.}]
			\step[fieldsource=\regexp{$MAPLOOP}, match={Radiation}, replace={Radiat.}]
			\step[fieldsource=\regexp{$MAPLOOP}, match={Polytechnic}, replace={Polytech.}]
			\step[fieldsource=\regexp{$MAPLOOP}, match={Radiology}, replace={Radiol.}]
			\step[fieldsource=\regexp{$MAPLOOP}, match={Practice}, replace={Pract.}]
			\step[fieldsource=\regexp{$MAPLOOP}, match={Reactor}, replace={React.}]
			\step[fieldsource=\regexp{$MAPLOOP}, match={Precision}, replace={Precis.}]
			\step[fieldsource=\regexp{$MAPLOOP}, match={Receivers}, replace={Receiv.}]
			\step[fieldsource=\regexp{$MAPLOOP}, match={Principles}, replace={Princ.}]
			\step[fieldsource=\regexp{$MAPLOOP}, match={Recognition}, replace={Recognit.}]
			\step[fieldsource=\regexp{$MAPLOOP}, match={Proceedings}, replace={Proc.}]
			\step[fieldsource=\regexp{$MAPLOOP}, match={Record}, replace={Rec.}]
			\step[fieldsource=\regexp{$MAPLOOP}, match={Processing}, replace={Process.}]
			\step[fieldsource=\regexp{$MAPLOOP}, match={Rehabilitation}, replace={Rehabil.}]
			\step[fieldsource=\regexp{$MAPLOOP}, match={Conversion}, replace={Convers.}]
			\step[fieldsource=\regexp{$MAPLOOP}, match={Digital}, replace={Digit.}]
			\step[fieldsource=\regexp{$MAPLOOP}, match={Convention}, replace={Conv.}]
			\step[fieldsource=\regexp{$MAPLOOP}, match={Disclosure}, replace={Discl.}]
			\step[fieldsource=\regexp{$MAPLOOP}, match={Correspondence}, replace={Corresp.}]
			\step[fieldsource=\regexp{$MAPLOOP}, match={Discussions}, replace={Discuss.}]
			\step[fieldsource=\regexp{$MAPLOOP}, match={Critical}, replace={Crit.}]
			\step[fieldsource=\regexp{$MAPLOOP}, match={Dissertations}, replace={Diss.}]
			\step[fieldsource=\regexp{$MAPLOOP}, match={Crystal}, replace={Cryst.}]
			\step[fieldsource=\regexp{$MAPLOOP}, match={Distributed}, replace={Distrib.}]
			\step[fieldsource=\regexp{$MAPLOOP}, match={Crystallography}, replace={Crystallogr.}]
			\step[fieldsource=\regexp{$MAPLOOP}, match={Dynamics}, replace={Dyn.}]
			\step[fieldsource=\regexp{$MAPLOOP}, match={Cybernetics}, replace={Cybern.}]
			\step[fieldsource=\regexp{$MAPLOOP}, match={Earthquake}, replace={Earthq.}]
			\step[fieldsource=\regexp{$MAPLOOP}, match={Decision}, replace={Decis.}]
			\step[fieldsource=\regexp{$MAPLOOP}, match={Economics}, replace={Econ.}]
			\step[fieldsource=\regexp{$MAPLOOP}, match={Economic}, replace={Econ.}]
			\step[fieldsource=\regexp{$MAPLOOP}, match={Economical}, replace={Econ.}]
			\step[fieldsource=\regexp{$MAPLOOP}, match={Edition}, replace={Ed.}]
			\step[fieldsource=\regexp{$MAPLOOP}, match={Evolutionary}, replace={Evol.}]
			\step[fieldsource=\regexp{$MAPLOOP}, match={Education}, replace={Educ.}]
			\step[fieldsource=\regexp{$MAPLOOP}, match={Exhibition}, replace={Exhib.}]
			\step[fieldsource=\regexp{$MAPLOOP}, match={Electrical}, replace={Elect.}]
			\step[fieldsource=\regexp{$MAPLOOP}, match={Electric}, replace={Elect.}]
			\step[fieldsource=\regexp{$MAPLOOP}, match={Experimental}, replace={Exp.}]
			\step[fieldsource=\regexp{$MAPLOOP}, match={Electrification}, replace={Electrific.}]
			\step[fieldsource=\regexp{$MAPLOOP}, match={Exploratory}, replace={Explor.}]
			\step[fieldsource=\regexp{$MAPLOOP}, match={Electromagnetic}, replace={Electromagn.}]
			\step[fieldsource=\regexp{$MAPLOOP}, match={Exposition}, replace={Expo.}]
			\step[fieldsource=\regexp{$MAPLOOP}, match={Electroacoustic}, replace={Electroacoust.}]
			\step[fieldsource=\regexp{$MAPLOOP}, match={Express}, replace={Express}]
			\step[fieldsource=\regexp{$MAPLOOP}, match={Electronics}, replace={Electron.}]
			\step[fieldsource=\regexp{$MAPLOOP}, match={Electronic}, replace={Electron.}]
			\step[fieldsource=\regexp{$MAPLOOP}, match={Fabrication}, replace={Fabr.}]
			\step[fieldsource=\regexp{$MAPLOOP}, match={Emerging}, replace={Emerg.}]
			\step[fieldsource=\regexp{$MAPLOOP}, match={Faculty}, replace={Fac.}]
			\step[fieldsource=\regexp{$MAPLOOP}, match={Engineering}, replace={Eng.}]
			\step[fieldsource=\regexp{$MAPLOOP}, match={Engineers}, replace={Eng.}]
			\step[fieldsource=\regexp{$MAPLOOP}, match={Engineer}, replace={Eng.}]
			\step[fieldsource=\regexp{$MAPLOOP}, match={Ferroelectrics}, replace={Ferroelect.}]
			\step[fieldsource=\regexp{$MAPLOOP}, match={Environment}, replace={Environ.}]
			\step[fieldsource=\regexp{$MAPLOOP}, match={Francais, French}, replace={Fr.}]
			\step[fieldsource=\regexp{$MAPLOOP}, match={Equations}, replace={Equ.}]
			\step[fieldsource=\regexp{$MAPLOOP}, match={Frequency}, replace={Freq.}]
			\step[fieldsource=\regexp{$MAPLOOP}, match={Equipment}, replace={Equip.}]
			\step[fieldsource=\regexp{$MAPLOOP}, match={Foundation}, replace={Found.}]
			\step[fieldsource=\regexp{$MAPLOOP}, match={Ergonomics}, replace={Ergonom.}]
			\step[fieldsource=\regexp{$MAPLOOP}, match={Fundamental}, replace={Fundam.}]
			\step[fieldsource=\regexp{$MAPLOOP}, match={European}, replace={Eur.}]
			\step[fieldsource=\regexp{$MAPLOOP}, match={Generation}, replace={Gener.}]
			\step[fieldsource=\regexp{$MAPLOOP}, match={Evaluation}, replace={Eval.}]
			\step[fieldsource=\regexp{$MAPLOOP}, match={Geology}, replace={Geol.}]
			\step[fieldsource=\regexp{$MAPLOOP}, match={Geophysics}, replace={Geophys.}]
			\step[fieldsource=\regexp{$MAPLOOP}, match={Innovations}, replace={Innov.}]
			\step[fieldsource=\regexp{$MAPLOOP}, match={Innovation}, replace={Innov.}]
			\step[fieldsource=\regexp{$MAPLOOP}, match={Geoscience}, replace={Geosci.}]
			\step[fieldsource=\regexp{$MAPLOOP}, match={Institute}, replace={Inst.}]
			\step[fieldsource=\regexp{$MAPLOOP}, match={Institution}, replace={Inst.}]
			\step[fieldsource=\regexp{$MAPLOOP}, match={Graphics}, replace={Graph.}]
			\step[fieldsource=\regexp{$MAPLOOP}, match={Instrument}, replace={Instrum.}]
			\step[fieldsource=\regexp{$MAPLOOP}, match={Guidance}, replace={Guid.}]
			\step[fieldsource=\regexp{$MAPLOOP}, match={Instrumentation}, replace={Instrum.}]
			\step[fieldsource=\regexp{$MAPLOOP}, match={Harmonics}, replace={Harmon.}]
			\step[fieldsource=\regexp{$MAPLOOP}, match={Harmonic}, replace={Harmon.}]
			\step[fieldsource=\regexp{$MAPLOOP}, match={Insulation}, replace={Insul.}]
			\step[fieldsource=\regexp{$MAPLOOP}, match={History}, replace={Hist.}]
			\step[fieldsource=\regexp{$MAPLOOP}, match={Integrated}, replace={Integr.}]
			\step[fieldsource=\regexp{$MAPLOOP}, match={Horizon}, replace={Horiz.}]
			\step[fieldsource=\regexp{$MAPLOOP}, match={Intelligence}, replace={Intell.}]
			\step[fieldsource=\regexp{$MAPLOOP}, match={Hungary}, replace={Hung.}]
			\step[fieldsource=\regexp{$MAPLOOP}, match={Hungarian}, replace={Hung.}]
			\step[fieldsource=\regexp{$MAPLOOP}, match={Intelligent}, replace={Intell.}]
			\step[fieldsource=\regexp{$MAPLOOP}, match={Hydraulics}, replace={Hydraul.}]
			\step[fieldsource=\regexp{$MAPLOOP}, match={Interactions}, replace={Interact.}]
			\step[fieldsource=\regexp{$MAPLOOP}, match={Hydrology}, replace={Hydrol.}]
			\step[fieldsource=\regexp{$MAPLOOP}, match={Internationales}, replace={Int.}]
			\step[fieldsource=\regexp{$MAPLOOP}, match={International}, replace={Int.}]
			\step[fieldsource=\regexp{$MAPLOOP}, match={Illuminating}, replace={Illum.}]
			\step[fieldsource=\regexp{$MAPLOOP}, match={Isotopes}, replace={Isot.}]
			\step[fieldsource=\regexp{$MAPLOOP}, match={Imaging}, replace={Imag.}]
			\step[fieldsource=\regexp{$MAPLOOP}, match={Israel}, replace={Isr.}]
			\step[fieldsource=\regexp{$MAPLOOP}, match={Industrial}, replace={Ind.}]
			\step[fieldsource=\regexp{$MAPLOOP}, match={Japan}, replace={Jpn.}]
			\step[fieldsource=\regexp{$MAPLOOP}, match={Information}, replace={Inf.}]
			\step[fieldsource=\regexp{$MAPLOOP}, match={Journal}, replace={J.}]
			\step[fieldsource=\regexp{$MAPLOOP}, match={Informatics}, replace={Inform.}]
			\step[fieldsource=\regexp{$MAPLOOP}, match={Knowledge}, replace={Knowl.}]
			\step[fieldsource=\regexp{$MAPLOOP}, match={Laboratory}, replace={Lab.}]
			\step[fieldsource=\regexp{$MAPLOOP}, match={Laboratories}, replace={Lab.}]
			\step[fieldsource=\regexp{$MAPLOOP}, match={Mathematical}, replace={Math.}]
			\step[fieldsource=\regexp{$MAPLOOP}, match={Language}, replace={Lang.}]
			\step[fieldsource=\regexp{$MAPLOOP}, match={Mathematics}, replace={Math.}]
			\step[fieldsource=\regexp{$MAPLOOP}, match={Abstracts}, replace={Abstr.}]
			\step[fieldsource=\regexp{$MAPLOOP}, match={Analysis}, replace={Anal.}]
			\step[fieldsource=\regexp{$MAPLOOP}, match={Academy}, replace={Acad.}]
			\step[fieldsource=\regexp{$MAPLOOP}, match={Annals}, replace={Ann.}]
			\step[fieldsource=\regexp{$MAPLOOP}, match={Accelerator}, replace={Accel.}]
			\step[fieldsource=\regexp{$MAPLOOP}, match={Annual}, replace={Annu.}]
			\step[fieldsource=\regexp{$MAPLOOP}, match={Acoustics}, replace={Acoust.}]
			\step[fieldsource=\regexp{$MAPLOOP}, match={Apparatus}, replace={App.}]
			\step[fieldsource=\regexp{$MAPLOOP}, match={Active}, replace={Act.}]
			\step[fieldsource=\regexp{$MAPLOOP}, match={Applications}, replace={Appl.}]
			\step[fieldsource=\regexp{$MAPLOOP}, match={Administration}, replace={Admin.}]
			\step[fieldsource=\regexp{$MAPLOOP}, match={Applied}, replace={Appl.}]
			\step[fieldsource=\regexp{$MAPLOOP}, match={Administrative}, replace={Administ.}]
			\step[fieldsource=\regexp{$MAPLOOP}, match={Approximate}, replace={Approx.}]
			\step[fieldsource=\regexp{$MAPLOOP}, match={Advanced}, replace={Adv.}]
			\step[fieldsource=\regexp{$MAPLOOP}, match={Advances}, replace={Adv.}]
			\step[fieldsource=\regexp{$MAPLOOP}, match={Archives}, replace={Arch.}]
			\step[fieldsource=\regexp{$MAPLOOP}, match={Archive}, replace={Arch.}]
			\step[fieldsource=\regexp{$MAPLOOP}, match={Aeronautics}, replace={Aeronaut.}]
			\step[fieldsource=\regexp{$MAPLOOP}, match={Artificial}, replace={Artif.}]
			\step[fieldsource=\regexp{$MAPLOOP}, match={Aerospace}, replace={Aerosp.}]
			\step[fieldsource=\regexp{$MAPLOOP}, match={Assembly}, replace={Assem.}]
			\step[fieldsource=\regexp{$MAPLOOP}, match={Affective}, replace={Affect.}]
			\step[fieldsource=\regexp{$MAPLOOP}, match={Association}, replace={Assoc.}]
			\step[fieldsource=\regexp{$MAPLOOP}, match={Africa}, replace={Afr.}]
			\step[fieldsource=\regexp{$MAPLOOP}, match={African}, replace={Afr.}]
			\step[fieldsource=\regexp{$MAPLOOP}, match={Astronomy}, replace={Astron.}]
			\step[fieldsource=\regexp{$MAPLOOP}, match={Aircraft}, replace={Aircr.}]
			\step[fieldsource=\regexp{$MAPLOOP}, match={Astronautics}, replace={Astronaut.}]
			\step[fieldsource=\regexp{$MAPLOOP}, match={Algebraic}, replace={Algebr.}]
			\step[fieldsource=\regexp{$MAPLOOP}, match={Astrophysics}, replace={Astrophys.}]
			\step[fieldsource=\regexp{$MAPLOOP}, match={American}, replace={Amer.}]
			\step[fieldsource=\regexp{$MAPLOOP}, match={Atmosphere}, replace={Atmos.}]
			\step[fieldsource=\regexp{$MAPLOOP}, match={Atomic}, replace={At.}]
			\step[fieldsource=\regexp{$MAPLOOP}, match={Atoms}, replace={At.}]
			\step[fieldsource=\regexp{$MAPLOOP}, match={Broadcasting}, replace={Broadcast.}]
			\step[fieldsource=\regexp{$MAPLOOP}, match={Australasian}, replace={Australas.}]
			\step[fieldsource=\regexp{$MAPLOOP}, match={Bulletin}, replace={Bull.}]
			\step[fieldsource=\regexp{$MAPLOOP}, match={Australia}, replace={Aust.}]
			\step[fieldsource=\regexp{$MAPLOOP}, match={Bureau}, replace={Bur.}]
			\step[fieldsource=\regexp{$MAPLOOP}, match={{Automatic }}, replace={Autom.}]
			\step[fieldsource=\regexp{$MAPLOOP}, match={Business}, replace={Bus.}]
			\step[fieldsource=\regexp{$MAPLOOP}, match={Automation}, replace={Automat.}]
			\step[fieldsource=\regexp{$MAPLOOP}, match={Canadian}, replace={Can.}]
			\step[fieldsource=\regexp{$MAPLOOP}, match={Automotive}, replace={Automot.}]
			\step[fieldsource=\regexp{$MAPLOOP}, match={Automobiles}, replace={Automob.}]
			\step[fieldsource=\regexp{$MAPLOOP}, match={Automobile}, replace={Automob.}]
			\step[fieldsource=\regexp{$MAPLOOP}, match={Ceramic}, replace={Ceram.}]
			\step[fieldsource=\regexp{$MAPLOOP}, match={Chemical}, replace={Chem.}]
			\step[fieldsource=\regexp{$MAPLOOP}, match={Behavioral}, replace={Behav.}]
			\step[fieldsource=\regexp{$MAPLOOP}, match={Behavior}, replace={Behav.}]
			\step[fieldsource=\regexp{$MAPLOOP}, match={Chinese}, replace={Chin.}]
			\step[fieldsource=\regexp{$MAPLOOP}, match={Belgian}, replace={Belg.}]
			\step[fieldsource=\regexp{$MAPLOOP}, match={Climatology}, replace={Climatol.}]
			\step[fieldsource=\regexp{$MAPLOOP}, match={Biochemical}, replace={Biochem.}]
			\step[fieldsource=\regexp{$MAPLOOP}, match={Clinical}, replace={Clin.}]
			\step[fieldsource=\regexp{$MAPLOOP}, match={Bioinformatics}, replace={Bioinf.}]
			\step[fieldsource=\regexp{$MAPLOOP}, match={Cognitive}, replace={Cogn.}]
			\step[fieldsource=\regexp{$MAPLOOP}, match={Biology, Biological}, replace={Biol.}]
			\step[fieldsource=\regexp{$MAPLOOP}, match={Colloquium}, replace={Colloq.}]
			\step[fieldsource=\regexp{$MAPLOOP}, match={Kolloquium}, replace={Kolloq.}]
			\step[fieldsource=\regexp{$MAPLOOP}, match={Biomedical}, replace={Biomed.}]
			\step[fieldsource=\regexp{$MAPLOOP}, match={Communications}, replace={Commun.}]
			\step[fieldsource=\regexp{$MAPLOOP}, match={Communication}, replace={Commun.}]
			\step[fieldsource=\regexp{$MAPLOOP}, match={Biophysics}, replace={Biophys.}]
			\step[fieldsource=\regexp{$MAPLOOP}, match={Compatibility}, replace={Compat.}]
			\step[fieldsource=\regexp{$MAPLOOP}, match={British}, replace={Brit.}]
			\step[fieldsource=\regexp{$MAPLOOP}, match={Components}, replace={Compon.}]
			\step[fieldsource=\regexp{$MAPLOOP}, match={Component}, replace={Compon.}]
			\step[fieldsource=\regexp{$MAPLOOP}, match={Computational}, replace={Comput.}]
			\step[fieldsource=\regexp{$MAPLOOP}, match={Delivery}, replace={Del.}]
			\step[fieldsource=\regexp{$MAPLOOP}, match={Computers}, replace={Comput.}]
			\step[fieldsource=\regexp{$MAPLOOP}, match={Computer}, replace={Comput.}]
			\step[fieldsource=\regexp{$MAPLOOP}, match={Department}, replace={Dept.}]
			\step[fieldsource=\regexp{$MAPLOOP}, match={Computing}, replace={Comput.}]
			\step[fieldsource=\regexp{$MAPLOOP}, match={Design}, replace={Des.}]
			\step[fieldsource=\regexp{$MAPLOOP}, match={Condensed}, replace={Condens.}]
			\step[fieldsource=\regexp{$MAPLOOP}, match={Detector}, replace={Detect.}]
			\step[fieldsource=\regexp{$MAPLOOP}, match={Conferences}, replace={Conf.}]
			\step[fieldsource=\regexp{$MAPLOOP}, match={Conference}, replace={Conf.}]
			\step[fieldsource=\regexp{$MAPLOOP}, match={Development}, replace={Develop.}]
			\step[fieldsource=\regexp{$MAPLOOP}, match={Congress}, replace={Congr.}]
			\step[fieldsource=\regexp{$MAPLOOP}, match={Differential}, replace={Differ.}]
			\step[fieldsource=\regexp{$MAPLOOP}, match={Consumer}, replace={Consum.}]
			\step[fieldsource=\regexp{$MAPLOOP}, match={Digest}, replace={Dig.}] 
			\step[fieldsource=\regexp{$MAPLOOP}, match={{ of the }}, replace={{ }}]
			\step[fieldsource=\regexp{$MAPLOOP}, match={{ of }}, replace={{ }}]
			\step[fieldsource=\regexp{$MAPLOOP}, match={{Of }}, replace={{ }}]
			\step[fieldsource=\regexp{$MAPLOOP}, match={{ on }}, replace={{ }}]
			\step[fieldsource=\regexp{$MAPLOOP}, match={{ On }}, replace={{ }}]
			\step[fieldsource=\regexp{$MAPLOOP}, match={{On }}, replace={{ }}]
			\step[fieldsource=\regexp{$MAPLOOP}, match={{ in }}, replace={{ }}]
			\step[fieldsource=\regexp{$MAPLOOP}, match=\regexp{\\\x{26}}, replace={{and}}] 
			\step[fieldsource=\regexp{$MAPLOOP}, match={{, and }}, replace={{, }}]
			\step[fieldsource=\regexp{$MAPLOOP}, match={{, And }}, replace={{, }}]
			\step[fieldsource=\regexp{$MAPLOOP}, match={{ and }}, replace={{ }}]
			\step[fieldsource=\regexp{$MAPLOOP}, match={{ And }}, replace={{ }}]
			\step[fieldsource=\regexp{$MAPLOOP}, match={{ In }}, replace={{ }}]
			\step[fieldsource=\regexp{$MAPLOOP}, match={{In }}, replace={{ }}]
			\step[fieldsource=\regexp{$MAPLOOP}, match={{ the }}, replace={{ }}] 
			\step[fieldsource=\regexp{$MAPLOOP}, match={{ The }}, replace={{ }}] 
			\step[fieldsource=\regexp{$MAPLOOP}, match={{The }}, replace={}] 
			\step[fieldsource=\regexp{$MAPLOOP}, match={{ for }}, replace={{ }}] 
			\step[fieldsource=\regexp{$MAPLOOP}, match={First}, replace={1st}]
			\step[fieldsource=\regexp{$MAPLOOP}, match={Second}, replace={2nd}]
			\step[fieldsource=\regexp{$MAPLOOP}, match={Third}, replace={3rd}]
			\step[fieldsource=\regexp{$MAPLOOP}, match={Fourth}, replace={4th}]
			\step[fieldsource=\regexp{$MAPLOOP}, match={Fifth}, replace={5th}]
			\step[fieldsource=\regexp{$MAPLOOP}, match={Sixth}, replace={6th}]
			\step[fieldsource=\regexp{$MAPLOOP}, match={Seventh}, replace={7th}]
			\step[fieldsource=\regexp{$MAPLOOP}, match={Eighth}, replace={8th}]
			\step[fieldsource=\regexp{$MAPLOOP}, match={Ninth}, replace={9th}]
			\step[fieldsource=\regexp{$MAPLOOP}, match={Tenth}, replace={10th}]
			\step[fieldsource=\regexp{$MAPLOOP}, match={Eleventh}, replace={11th}]
			\step[fieldsource=\regexp{$MAPLOOP}, match={Twelfth}, replace={12th}]
			\step[fieldsource=\regexp{$MAPLOOP}, match={Thirteenth}, replace={13th}]
			\step[fieldsource=\regexp{$MAPLOOP}, match={Fourteenth}, replace={14th}]
			\step[fieldsource=\regexp{$MAPLOOP}, match={Fifteenth}, replace={15th}]
			\step[fieldsource=\regexp{$MAPLOOP}, match={Sixteenth}, replace={16th}]
			\step[fieldsource=\regexp{$MAPLOOP}, match={Seventeenth}, replace={17th}]
			\step[fieldsource=\regexp{$MAPLOOP}, match={Eighteenth}, replace={18th}]
			\step[fieldsource=\regexp{$MAPLOOP}, match={Nineteenth}, replace={19th}]
			\step[fieldsource=\regexp{$MAPLOOP}, match={Twentieth}, replace={20th}]
		}            
    }
}%

\makeatletter
\newcommand{\linebreakand}{%
  \end{@IEEEauthorhalign}
  \hfill\mbox{}\par
  \mbox{}\hfill\begin{@IEEEauthorhalign}
}
\makeatother

\title{What's Really Different with AI? -- \\A Behavior-based Perspective on System Safety for Automated Driving Systems}

\author{\IEEEauthorblockN{%
    Marcus Nolte\IEEEauthorrefmark{1}\orcidlink{0000-0003-4909-9403},
    Nayel Fabian Salem\IEEEauthorrefmark{1}\orcidlink{0000-0003-4909-9403},
    Olaf~Franke\IEEEauthorrefmark{2},
    Jan~Heckmann\IEEEauthorrefmark{3},
    Christoph Höhmann\IEEEauthorrefmark{4},\\
    Georg Stettinger\IEEEauthorrefmark{5}\orcidlink{0000-0001-5607-3557},
    Markus Maurer\IEEEauthorrefmark{1}\orcidlink{0000-0002-5357-9701}
}\\
\linebreakand
\IEEEauthorblockA{%
    \IEEEauthorrefmark{1}\textit{TU Braunschweig} \\%
    \textit{Institute of Control Engineering}, Braunschweig, Germany\\%
    \{m.nolte, n.salem, m.maurer\}@tu-braunschweig.de%
}
\and
\IEEEauthorblockA{%
    \IEEEauthorrefmark{2}\textit{MAN Truck \& Bus SE}, Munich, Germany\\%
    olaf.franke@man.eu\\%
}%
\linebreakand%
\IEEEauthorblockA{%
    \IEEEauthorrefmark{3}%
    \textit{Deutsche Bahn Regio AG}, Berlin, Germany\\%
    jan.heckmann@deutschebahn.com%
}%
\and%
\IEEEauthorblockA{%
    \IEEEauthorrefmark{4}%
    \textit{Mercedes-Benz AG}, Stuttgart, Germany\\%
    christoph.hoehmann@mercedes-benz.de%
}
\linebreakand
\IEEEauthorblockA{%
    \IEEEauthorrefmark{5}%
    \textit{Infineon Technologies AG}, Neubiberg, Germany\\%
    georg.stettinger@infineon.com%
}%
\thanks{M. Nolte and N. F. Salem contributed equally to this work.}
}%
\begin{document}%
\thispagestyle{empty}
\twocolumn[
\begin{@twocolumnfalse}
	\large {\copyright\ 2025 IEEE. Personal use of this material is permitted. Permission from IEEE must be obtained for all other uses, in any current or future media, including reprinting/republishing this material for advertising or promotional purposes, creating new collective works, for resale or redistribution to servers or lists, or reuse of any copyrighted component of this work in other works.} \\ \\
	
	{\Large Accepted to be published in \emph{2025 IEEE International Automated Vehicle Validation Conference (IAVVC)}, Baden-Baden, Germany, 2025.} \\ \\
 
	
	Cite as:
	\vspace{0.1cm}
	
	\noindent\fbox{%
		\begin{minipage}{0.98\textwidth}%
			M.~Nolte, N.~F.~Salem, O.~Franke, J.~Heckmann, C.~Höhmann, G.~Stettinger, and M.~Maurer, ``{What's} {Really} {Different} with {AI} ? -- {A} {Behavior-based} {Perspective} on {System} {Safety} for {Automated} {Driving} {Systems},'' in \emph{2025 IEEE International Automated Vehicle Validation Conference}, to be published.
		\end{minipage}
	}
	\vspace{2cm}
	
\end{@twocolumnfalse}
]

\noindent%
\hologo{BibTeX}:

\noindent
	\begin{centering}
	\footnotesize
	\begin{lstlisting}[frame=single,linewidth=\textwidth]
@inproceedings{nolte_salem_AI_safety_2025,
    author = {Nolte, Marcus and Salem, Nayel Fabian and Franke, Olaf and Heckmann, Jan and H\"ohmann, Christoph and Stettinger, Georg and Maurer, Markus},
    booktitle = {12},
    title = {{What's} {Really} {Different} with {AI} ? -- {A} {Behavior-based} {Perspective} on {System} {Safety} for {Automated} {Driving} {Systems}},
    address = {Baden-Baden, Germany},
    year = {2025},
    publisher = {IEEE. to be published},
}
	\end{lstlisting}
\end{centering}
\maketitle%
\thispagestyle{empty}%
\pagestyle{empty}%
%
%

\begin{abstract}
Assuring safety for ``AI-based'' systems is one of the current challenges in safety engineering.
For automated driving systems, in particular, further assurance challenges result from the open context that the systems need to operate in after deployment.
The current standardization and regulation landscape for ``AI-based'' systems is becoming ever more complex, as standards and regulations are being released at high frequencies.

This position paper seeks to provide guidance for making qualified arguments which standards should meaningfully be applied to (``AI-based'') automated driving systems.
Furthermore, we argue for clearly differentiating sources of risk between AI-specific and general uncertainties related to the open context.
In our view, a clear conceptual separation can help to exploit commonalities that can close the gap between system-level and AI-specific safety analyses, while ensuring the required rigor for engineering safe ``AI-based'' systems.
\end{abstract}

%

\section{Introduction}
\label{sec:intro}

Alongside the ongoing deployment of Automated Driving Systems (ADS) and 
recent technological progress in Artificial Intelligence (AI), huge research efforts are currently geared toward safety assurance for “AI-based” automated driving systems. These efforts are sidelined by the release of a plethora of different standards and regulations for both, general AI-based applications and AI-based automated driving systems.

The fast-paced standardization and regulation landscape poses challenges to all stakeholders involved in the development of automated vehicles: Product-compliance laws typically demand a development aligned with the current state of the art. Depending on the actual regulation, this state of the art refers to combinations of research, standardization, industry best practices.
Consistent regulation and standardization is critical here to support a safe industrial realization of ADS, as contradicting definitions of what an ``AI-based system'' is or what ``safety'' means can lead to competitive disadvantages in more tightly regulated markets such as the European Union.

Three main challenges follow from this which are related to a) extracting relevant demands from closely related standards that consider ADS, ``AI-based'' systems or both, b) providing compelling arguments, why a standard might \emph{not} be applicable to the developed system, and c) defining processes that efficiently address relevant contributions to the state of the art from such standards, e.g. by establishing methods for safety analyses that allow addressing multiple standards at once. 
These tasks are further complicated by different definitions in current standardization and regulation for what constitutes an ``AI-based'' system.

In the context of safety assurance for ``AI-based'' automated driving systems, it is crucial to clearly differentiate possible sources of risk.
Previous work in this context (e.g. \cite{burton2023a, burton2023, nolte2024}, cf. also \cref{fig:uncertainty}) has argued the importance of tracing back relevant risks to different sources of uncertainty that affect a system of interest.
Such uncertainty can, e.g., include knowledge gaps related to the open world in which a system needs to operate or uncertainty stemming from the world's complexity (e.g. such as unpredictability of long-term interactions between traffic participants).
This type of uncertainty reflects in general challenges related to the definition of safe system-level behavior, regardless of how a system is implemented.
It will, however, also have an impact on specific ``AI-related'' challenges in the sense that this type of uncertainty requires addressing the representativeness of training data or assessing the generalization capabilities of data-driven algorithms in contexts that are unknown at design or training time.

Further uncertainty stems from the complexity of the system itself and the (``AI-'') algorithms that are applied to implement system functionality.
It becomes crucial to differentiate what risks are related to uncertainty caused by emergent system properties that are related to complex \emph{system architectures} (i.e., high numbers of interrelated system components) and what risks are related to the specific nature of e.g. the black-box properties of \emph{(network) architectures} in data-based algorithms.
To provide guidance in the current regulation and standardization landscape, this position paper will address the challenges related to differing ``AI'' definitions (\cref{sec:background}).

Furthermore, we find that current research, standards and regulation are separating assurance challenges too stringently into the categories of ``AI-related'' and ``non-AI-related''. 
Often, explicit differences between ``AI-based'' and ``classic'' systems \cite{iso8800, ashmore2021, schnitzer2024} are emphasized.\footnote{From a Safety Engineering perspective, systems that learn in an unsupervised fashion at runtime (``online methods'' as per ISO/PAS~8800) represent  a whole additional category of challenges. We will focus on systems trained offline for this paper. ISO/PAS~8800 explicitly states \cite[145]{iso8800} that online methods for retraining to e.g. learn distribution shift can be subject to ``other requirements''.}
We will compare challenges for the safety assurance of automated driving systems operating in open contexts with current challenges regarding safety assurance for ``AI-based'' systems.

Based on this comparison, we critically discuss the question: ``What's \emph{really} different with AI?'', when it comes to assessing causes of risk related to artificial intelligence and system safety for automated vehicles (\cref{sec:related_work}). 
We highlight the importance of providing answers to very specific questions, such as: \emph{What risks are specific to the implementation of AI algorithms?} \emph{How do these risks impact established assurance practices in the automotive industry?}
As a partial answer to these questions, we discuss behavior-based aspects of safety assurance that help establish engineering rigor for autonomous systems which operate in an open context (\cref{sec:behavior}).

The paper will conclude with a set of recommendations and a call to action when considering the identified commonalities regarding system safety for ``AI-based'' systems (\cref{sec:Conclusion}).
The recommendations are meant to provide a \emph{starting point} to establish rigorous traceability from system-level safety aspects to AI-specific safety aspects that is currently e.g. demanded by ISO/PAS~8800:2022 \cite{iso8800} without providing specific methodical guidance.
This paper will provide according high-level suggestions in \cref{sec:behavior}. 
Further research initiatives on their refinement and validation will still be required.
%

\section{The Troubles with AI Definitions}
\label{sec:background}
For dealing with AI-related assurance challenges, a clear definition of what is considered as an ``AI-based'' system is paramount.
Regarding regulation and standardization, varying definitions may influence the scope of regulations and standards.
The differing notions of \emph{artificial intelligence} in current standards and regulation may hence pose hurdles when navigating the current standardization and regulation landscape.

The EU AI Act (Regulation (EU) 2024/1689, \cite{eu1689}) defines an \emph{AI System} as \enquote{[...] a machine-based system that is designed to operate with varying levels of autonomy and that may exhibit adaptiveness after deployment, and that, for explicit or implicit objectives, infers, from the input it receives, how to generate outputs such as predictions, content, recommendations, or decisions that can influence physical or virtual environments.}
This is an extremely broad and partially blurry definition:
The degree of autonomy or adaptiveness remain undefined, while any sufficiently complex rule-based system can be considered to infer predictions or make decisions based on its inputs.

Similar, broad definitions are adopted by several ISO standards related to ``AI Systems'': 
ISO/IEC~22989 \cite{iso22989} follows similar concepts as the EU AI Act.
The standard defines an ``AI system'' as an \enquote{engineered system that generates outputs such as content, forecasts, recommendations or decisions for a given set of human-defined objectives}\cite[Def.~3.7]{iso22989}.
ISO/IEC~5338 \cite{iso5338} and ISO/IEC~23894 \cite{iso23894} follow the same definition.
Similar challenges as for the EU Act definition arise here, as ``human-defined objectives'' extend the definition to any type of system that follows human-defined rulesets.

ISO/IEC~TR~24028 \cite{iso24028} and ISO~29119 \cite{iso29119} define artificial intelligence in a slightly more narrow fashion as the \enquote{capability of an engineered system to acquire, process and apply knowledge and skills}\cite[Def.~3.4]{iso24028}, with an \emph{engineered system} being a \enquote{combination of interacting elements organized to achieve one or more stated purposes} \cite[Def.~3.38]{iso24028}. \emph{Knowledge} here refers to \enquote{facts, information [...] and skills acquired through experience or education}\cite[Def.~3.4, NOTE~1]{iso24028}.
This is, again, problematic, as a) a definition of \emph{skills} is not given, and b) the applied knowledge definition attributes human-like properties to a technical system.

ISO/PAS~8800 \cite{iso8800}, which is of direct relevance for the design and assurance of automated driving systems in the context of ``AI-based systems", adopts a restricted definition, as it defines an ``AI system'' as an \enquote{item or element that utilises one or more AI models} \cite[Def.~3.1.17]{iso8800}, where an AI model is a \enquote{construct containing logical operations, arithmetical operations or a combination of both to generate an inference or prediction based on input data or information without being completely defined by human knowledge} \cite[Def.~3.1.7]{iso8800}.

The idea of defining an AI model as a construct that can infer results without being completely defined by human knowledge is helpful in so far as this definition provides a clear differentiating property: 
This comprises all data-driven (or ``machine-learning-based'') algorithms, but also more old-fashioned expert systems which may perform logical inference on ontological knowledge-bases, while every system that generates predictions, data or information on a mere set of pre-defined rules would not fall under the given definition.
For the remainder of the paper, we will use this definition to differentiate between ``AI-based'' (as in the definition of ISO/PAS~8800) and ``classic'' systems.

Given the challenges that come with the above-mentioned broad and differing definitions for ``AI-based'' systems, in the following, we will summarize related work that can help to scope assurance-related challenges for automated driving systems that contain AI models in the sense of ISO/PAS~8800. 
%

\section{Mapping Out Assurance Challenges}
\label{sec:related_work}
In the following, we will provide a literature overview in three categories: First, we will give an overview of assurance challenges that are related to risks emerging from the fact that automated vehicles are complex systems exposed to an open context.
Second, we will compare and contrast the collected challenges with (alleged) challenges that can be found in selected state-of the art approaches for the assurance of AI-based systems.
Finally, we will provide additional context for the arguments in \cref{sec:behavior} by addressing particular (safety) assurance challenges related to defining safe system behavior that are relevant in all aforementioned cases.

\subsection{Selected Assurance Challenges related to an Open System Context}
\label{sec:open-context}

\begin{figure*}[!htp]%
    \savebox{\largestimage}{\tikzset{%
	small dot/.style = {fill, circle, minimum width=.5em, inner sep=0pt},%
	DLbox/.style={
		shadow,%
		align=center,%
		shadedGray,%
		minimum width=4.15em,%
		inner ysep=0.25em%
	}%
}%
\begin{tikzpicture}[every node/.style={font=\footnotesize}]%
	\node[circle, shadow, minimum width=12.2em, shadedGrayMediumDark, label={[anchor=north, yshift=-0.21em,align=center]north:Input space\\ \& task}] (out) {};%
	\node[circle, shadow, minimum width=7.8em, shadedGrayMedium,label={[align=center,anchor=north, yshift=-0.5em]north:Data} ] (mid) {};%
	\node[circle, shadow, inner sep=0, minimum width=3.75em, shadedGrayMediumLight, align=center] (center) {ML \\Model};%
	\node[small dot] at (out.west) (v11) {};%
	\node[small dot] at (mid.west) (v2) {};%
	\node[small dot] at (center.west) (v3) {};%
	\draw[thick, -{Stealth[round, length=8pt]}] (v11) -- (v2);%
	\draw[thick, -{Stealth[round, length=8pt]}] (v2) -- (v3);%
	\node[align=center, anchor=east, left=1.3em of out.west] (complexity) {Environmental,\\ task, and \\system \\complexity};%
	\coordinate (c) at ($(out.east) + (0.2,0)$);%
	\draw[thick, decorate, decoration = {calligraphic brace, raise=-0em, amplitude=0.4em, mirror}] let \p1=(c), \p2=(center.south), \p3=(center.north) in (\x1, \y2) node (v1) {} -- node[xshift=0.1em, align=center, right]{Performance-\\insufficiency} (\x1, \y3) node (v2) {};%
	\draw[thick, decorate, decoration = {calligraphic brace, raise=-0em, amplitude=0.4em, mirror}] let \p1=(c), \p2=(center.north), \p3=(out.north) in (\x1, \y2) -- node[xshift=0.1em,align=center, right, name=test] {Specification-\\insufficiency} (\x1, \y3) node (v3) {};%
	\draw[densely dotted] (center.south) -- (v1.center);%
	\draw[densely dotted] (center.north) -- (v2.center);%
	\draw[densely dotted] (out.north) -- (v3.center);%
	\node[align=center, below left=1em and 0.5em of out.south] (a1) {Argument on sufficiency\\ of input space definition \\and requirements specification};%
	\node[align=center, below right=1em and 0.5em of out.south] (a2) {Argument on\\ sufficiency of datasets};%
	\node[align=center, below=4.5em of out.south] (a3) {Argument on performance\\ characterists of ML model};%
	\draw[line width=1em, tuLightBlue20, stroke arrow={ shorten back=0.1em, stroke color=tuBlue, stroke width=0.05em, arrow tip end=Triangle}] ($(complexity.east) - (0.25,0)$) -- (v11);%
	\draw[shorten >= .1em, thick] (a1) -- (out);%
	\draw[shorten >= .1em, thick] (a2) -- (mid);%
	\draw[shorten >= .1em, thick] (a3) -- (center);%
    \pgfresetboundingbox%
    \draw[draw=none, use as bounding box] let \p1=(a3.south), \p2=(complexity.west), \p3=(test.east), \p4=(out.north) in (\x2, \y1) rectangle (\x3, \y4 + 2);%
\end{tikzpicture}%
}%
    \begin{subcaptionblock}[t][][t]{.995\columnwidth}%
        \raisebox{\dimexpr\ht\largestimage-\height}{%
            \tikzset{%
		layer/.style n args={2}{%
		  rectangle,%
		  draw,%
		  fill,%
		  draw opacity=0.5,%
		  fill opacity=0.5,%
		  inner sep=0pt,%
		  fit=#1,%
		  label={[anchor=west, align=left, label distance=-0.5em, font=\sffamily]180:#2}}%
}%
\tikzset{%
	small dot/.style = {fill, circle, minimum width=.5em, inner sep=0pt}%
}%
\begin{tikzpicture}[every node/.style={font=\footnotesize}]%
	\node[circle, shadow, minimum width=13.5em, shadedGrayMediumDark, label={[anchor=north,align=center, yshift=-0.3em]north:Environment\\(World)}] (out) {};%
	\node[circle, shadow, minimum width=8.75em, shadedGrayMedium,label={[align=center,anchor=north, yshift=-0.4em]north:Observations\\(Evidence)} ] (mid) {};%
	\node[circle, shadow, inner sep=0, minimum width=3.75em, shadedGrayMediumLight, align=center] (center) {System\\(Decision\\maker)};%
	\node[small dot] at (out.west) (v1) {};%
	\node[small dot] at (mid.west) (v2) {};%
	\node[small dot] at (center.west) (v3) {};%
	\draw[thick, -{Stealth[round, length=8pt]}] (v1) -- (v2);%
	\draw[thick, -{Stealth[round, length=8pt]}] (v2) -- (v3);%
	\node[align=center, left=1.3em of out.west] (complexity) {Environmental,\\ task, and \\system \\complexity};%
	\node[align=center, right=2em of out.east] (risk) {Risk};%
	\node[align=center] (a3) at ($(out.south) - (0,0.5)$){Manifestations \\ of uncertainty};%
	\draw[line width=1em, tuLightBlue20, stroke arrow={ shorten back=0.1em, stroke color=tuBlue, stroke width=0.05em, arrow tip end=Triangle}]($(complexity.east) - (0.25,0)$) -- (v1);%
	\draw[line width=1em, tuLightBlue20, stroke arrow={ shorten back=0.1em, stroke color=tuBlue, stroke width=0.05em, arrow tip end=Triangle}] ($(out.east) + (0.15,0)$) -- (risk);%
    \pgfresetboundingbox%
    \draw[draw=none, use as bounding box] let \p1=(a3.south), \p2=(complexity.west), \p3=(risk.east), \p4=(out.north) in (\x2, \y1) rectangle (\x3, \y4 + 2);%
\end{tikzpicture}%
%
        }%
        \caption{General sources of uncertainty, related to an open world. Adapted from \cite{burton2023}: Modified by replacing assurance uncertainty with risk.\label{fig:general_uncertainty}}%
    \end{subcaptionblock}
    \hspace{\columnsep}
    \begin{subcaptionblock}[t][][t]{.995\columnwidth}%
        \usebox{\largestimage}%
        \caption{Specific sources of uncertainty that cause assurance challenges related to AI-based systems. Redrawn from \cite{burton2023}. \label{fig:ml_uncertainty}}
    \end{subcaptionblock}%
    \caption{Conceptual representations of uncertainty impacting systems (generic and AI-specific) operating in an open context according to \cite{burton2023}.\label{fig:uncertainty}}%
\end{figure*}%

As discussed in the introduction, assurance challenges related to automated driving systems are closely related to uncertainty that comes with the requirement to operate in an open world.
Assuring system properties (i.e., providing \enquote{grounds for justified confidence that a \emph{claim} [...] has been or will be satisfied} \cite[Def.~3.1.1]{iso15026-1}) such as safety becomes challenging: Not all situations that the engineered system will encounter over its lifetime can be foreseen by the developers at design time.

\citeauthor{burton2023} \cite{burton2023} give a visual explanation of uncertainty-related challenges (cf. \cref{fig:general_uncertainty}): A technical \emph{system} perceives the \emph{environment} through \emph{observations} and infers decisions or actions based on these observations.
The environment is subject to the aforementioned open-context-related uncertainty (e.g., unpredictability, incomplete knowledge about the world, inherent complexity).
The observations that a technical system makes are subject to technology-related uncertainty (e.g., measurement noise, limited fields of view, resolution limitations).
The system is subject to uncertainty, as it is complex in itself, must infer decisions based on incomplete information, and may apply non-deterministic algorithms. \cite[04]{burton2023}

Eventually, according to \citeauthor{burton2023} \cite{burton2023}, the aforementioned manifestations of uncertainty result in \emph{assurance uncertainty}, i.e., in uncertainty about the confidence that a certain claim has been or will be satisfied.
In the context of safety assurance (i.e., the process of assuring a claim about the \enquote{abscence of unreasonable risk} \cite[Part~1, Def.~3.132]{iso2018} for a system), \emph{risk} is strongly related to uncertainty (according to ISO~31000, risk is the \emph{effect of uncertainty on objectives} \cite[Def.~3.1]{iso31000}).
Hence, the assurance uncertainty according to \cite{burton2023} can be rephrased in terms of a residual risk (cf. \cref{fig:general_uncertainty}) that can be mitigated, but not fully eliminated, as at least the uncertainty related to the open world can never be eliminated.

For AV safety assurance, it must be proven that the residual risk is \emph{acceptable} or \emph{reasonable} with respect \enquote{to valid societal or moral concepts} \cite[Part~1, Def.~3.176]{iso2018}\footnote{This is a logical inversion of the definition for \emph{unreasonable} risk as per ISO~26262-1:2018 \cite[Def.~3.176]{iso2018}. Many technical standards such as ISO~26262 \cite{iso2018} interpret risk in relation to the occurrence of \emph{physical} harm. However, broader definitions of harm can include harm of stakeholder values \cite{nolte2024, salem2022}, including e.g. flawed interactions with emergency responders \cite{koopman2024a}.}.
Strategies that can be used to mitigate such residual risk below such a threshold must target the root causes of uncertainty.
These include methods for obtaining an understanding of the real world, e.g. by diligently analyzing technological requirements for data acquisition, or generating sufficient knowledge about the systems' internal and external behavior (cf. \cref{sec:behavior}).

Finally, the uncertainty related to the open context entails the need for diligent field monitoring \cite{burton2023a}. 
Due to the boundaries of human knowledge, developers cannot foresee every possible situation that a system can encounter in the field. 
If any evidence is found that development decisions do not hold and that this leads to an underestimation of the risk that the system poses to surrounding road users, this must trigger reevaluation processes and a rollout of required system updates. 
In severe cases, a good safety culture demands that these processes can ground the fleet until the issue is mitigated (cf. \cite[Sec.~8.3.4.1, pp. 286f.]{koopman2022}).

\subsection{Selected Assurance Challenges for AI-based Systems}
\label{sec:ai-assurance}
Assurance challenges for AI-based (as per ISO/PAS~8800) systems have been frequently discussed in the literature \cite{czarnecki2018, ashmore2021, burton2023, troubitsyna2024, stettinger2024}.
Main themes that can be identified in all related publications are: The \emph{black box} character of AI models when compared to classic model-based white or gray box systems, \emph{robustness} challenges regarding consistent outputs under minor input variations as well as \emph{explainability} challenges.

These challenges are not fully independent of each other, but are closely connected to the black box property mentioned above:
The AI models that constitute an AI-based system according to ISO~8800 are highly complex and nonlinear estimators, consisting of up to trillions of interacting parameters. 
This presents a degree of complexity that is, without applying proxy methods, not human comprehensible and turns the inner workings of those estimators into a factual black box for engineers.
This hinders the structural assurance of AI elements, compared to the assessment of system architectures, as it is the state of the art for "non-AI" system elements.
Further, the black box character leads to a lack of explainability \cite{ashmore2021, troubitsyna2024} regarding \emph{why} an AI model generates an output given a certain input.

Regarding robustness challenges, the high nonlinearity in the AI models makes up the core strength of machine learning approaches:
Given a suitable architecture, sufficient training and a sufficient number of parameters, any nonlinear input-output mapping can be approximated sufficiently close \cite{hornik1989}.
However, these nonlinearities can lead to significant changes in the output data (classifications, estimations, etc.) given only slight variations of the input data. This is the reason for the discussed robustness challenge \cite{troubitsyna2024} -- and presents the main challenge for assuring properties regarding the input-output relations of AI models.

In this respect, very specific research questions must be answered, to tackle the aforementioned challenges, such as:
``How can the translation from latent feature representations in black box models to explainable feature representations be improved and robustified?'' (cf, ``Explainable AI'' \cite{sokol2023}) or: ``How can the robustness of AI models be increased?''

\citeauthor{burton2023} \cite{burton2023} discuss different aspects of uncertainty which are closely related to AI component characteristics.
For this, the authors establish a corresponding model (\cref{fig:ml_uncertainty}) to the one which they use for discussing the impact of uncertainty on general complex autonomous systems (\cref{fig:general_uncertainty}).
According to \cite{burton2023}, the uncertainty related to the open world requires a diligent definition of the input space of AI models -- in our view, this requirement is no discriminator between AI-based systems and ``classic'' systems that must operate in an open context.

The inner rings in \cref{fig:ml_uncertainty} are, on the contrary, specific AI-related challenges:
While arguing safety over the sufficiency of datasets is closely related to a diligent understanding of the input space, properties of datasets, such as class labels, class distributions or coverage arguments on the sufficiency of data to cover the input space, are very specific to AI-based systems.

Further, \citeauthor{troubitsyna2024} \cite{troubitsyna2024} emphasize that safety is a system property.
The authors argue that there is an urgent need for research that relates performance metrics for AI models to system-level safety metrics.
Such a connection is required to enable assurance arguments on the performance characteristics of an ML model (inner circle in \cref{fig:ml_uncertainty}) in a system context.
This requires a close collaboration between safety and AI engineers.

Regarding the connection between system-level safety metrics and AI-specific performance metrics, ISO/PAS~8800 \cite{iso8800} assumes that safety analyses have been conducted and system-level requirements have been formulated, as a precondition for initiating the AI safety lifecycle.
This underpins that further work is needed to connect both levels of abstraction.

Finally, \citeauthor{troubitsyna2024} \cite{troubitsyna2024} argue that more engineering rigor must go into the design and assurance of AI-based systems.
ISO/PAS~8800 \cite{iso8800} aims to address this challenge by providing process guidelines, including processes for a diligent monitoring of AI performance over the entire system lifecycle.
This translates to life-cycle monitoring requirements, as discussed for the general operation of ADS in an open context (\cref{sec:open-context}).

\subsection{Intermediate Conclusion}
\label{sec:lit-summary}
\citeauthor{koopman2024} \cite[slide~6]{koopman2024} states that \enquote{Machine Learning breaks the Vee} in the sense that the application of data-based algorithms breaks engineering rigor.
Traditionally, engineering rigor allows building a strong prior (in a Bayesian sense) that a system is safe, which can be confirmed or refuted by evidence generated in the verification and validation process \cite{koopman2022}.

Considering the argumentation in this paper, it should have become clear that it is not only machine learning what breaks the established verification and validation concepts in rigorous engineering processes.
Major contributions to related assurance challenges stem from the uncertainty of an open system context.
This uncertainty is the root cause of assurance challenges that affect ADS in general.
In the Bayesian analogy given in \cite{koopman2024}, the challenge becomes to build well-formed and well-informed priors.
In consequence, this means that, regardless of the type of system, methods are required which help to reduce uncertainty where this is possible (e.g., through diligent analyses of the operational environment and expected ADS behavior), and which enable a diligent documentation of development assumptions with respect to the properties of the operational context.
Additionally, methods are required that allow to validate, verify, and, if required, update such assumptions over the entire system lifecycle \cite{burton2023a, troubitsyna2024}.

As ISO/PAS~8800 does explicitly not make the connection between system-level analyses or requirements and AI-specific analyses, further guidance is required for a targeted exploitation of commonalities and to make efficient connections between system-level and AI-specific analyses. 

\subsection{Selected Assurance Challenges for Safe System-Level Behavior}
\label{sec:behavioral-safety}
The challenge of connecting system-level metrics to AI-specific performance metrics is further complicated by the fact, that effective system-level metrics for ADS (safety) assurance are still an area of active research \cite{neurohr2021, chen2024, troubitsyna2024, galbas2024, fraade-blanar2025}.
Besides the \emph{definition} of these system-level metrics, their \emph{traceability} through the design and verification \& validation process remains a challenge.
Specifically, the meaningful decomposition of system-level metrics to V\&V pass-fail criteria remains challenging \cite{avsc2021, bsi1891, stettinger2024}.
Note that these challenges are fully independent of how the system is implemented.

Specific concepts that contribute to establishing traceability are discussed in both, the AI community \cite{barbier2024}, and the safety assurance community \cite{nolte2024, salem2024, bsi1891, galbas2024, stettinger2024}.
Related concepts are the \emph{Operational Design Domain} \cite{weissensteiner2023,barbier2024}, the required \emph{system behavior} \cite{ nolte2017, bsi1891, barbier2024, salem2024} as well as \emph{system capabilities} or \emph{behavioral competencies} \cite{nolte2017, nolte2024, avsc2021, stettinger2024}.

In the context of safety assurance, the definition of the Operational Design Domain, a corresponding behavior specification as well as corresponding sets of required behavioral competencies are artifacts that allow to capture key assumptions about how the automated vehicle interacts with the open world.
Such assumptions can be captured in terms of the elements which belong to the ODD, which characteristics of the ODD the vehicle needs to recognize, and how the vehicle should react to and interact with the ODD. 
A concept that allows to specify such interactions already at a very abstract level are sequences of \emph{maneuvers} \cite{bsi1891, salem2024, nolte2017}. 

The definition of the ODD, the respective desired behavior\footnote{ODD and behavior specification are highly iterative processes, as both can be mutual sources of constraints or extension needs.}, and the corresponding behavioral competencies comes with its own set of challenges:
A diligent definition of these artifacts requires a consequent derivation of behavioral requirements and relevant ODD elements from stakeholder needs.
Business goals, but also normative sources such as laws, standards or societal expectation are examples of origins of stakeholder needs.
For a traceable definition of artifacts and a documentation of justified assumptions, such normative sources, (semi-)formal representations of such sources are required \cite{salem2024}.
%

\section{Benefits of Behavior-Based Safety Analyses}
\label{sec:behavior}
The idea of making assumptions about the open context explicit by providing traceable definitions of the ODD, behavior, and behavioral competencies is a direct consequence of applying established systems engineering practice.
In the following, we will summarize corresponding concepts that can help to initiate the safety life cycle for automated driving systems.

\subsection{Systems Engineering Concepts: Operational Domain \& Operational Concept}
Modern Systems Engineering puts a particular focus on diligent problem space analyses.
The problem space is typically differentiated from the solution space.
The problem space is focused on understanding what properties (or capabilities) a system needs.
The solution space is focused on designing the system itself. 
In contrast to the solution domain, which would include technical and physical architectures and technology-specific solutions for realizing a system, it is good practice to keep problem-domain analyses solution and technology neutral.

From a Systems Engineering perspective, such considerations are summarized in the \emph{operational concept}, which describes system characteristics and how the system shall be operated from the perspective of relevant stakeholders.
As the focus is on system operation, i.e., its interaction with other entities and the world, the layer of abstraction that considers questions of stakeholder needs, use-cases, scenarios, system behavior, and required capabilities is called the \emph{operational domain}\footnote{As the term can be confused with the ADS Operational Design Domain, we will only refer to the \emph{operational concept} as an artifact in the following.}. \cite{walden2023}

\subsection{Mapping to Automated Driving Systems}
\label{sec:mapping}
These Systems Engineering concepts can help to provide additional structure to the assurance challenges related to automated driving systems.
When considering \cite{burton2023} and \mbox{ISO/PAS~8800}~\cite{iso8800}, all analyses related to defining an operational concept contribute to the definition of a systems' input space and task and capture assumptions regarding the open context. Table \ref{tbl:translation} provides a summary of the following steps.
\begin{table}
    \caption{Summary of terms used in \cref{sec:mapping}. Arrows should be read as a ``subsumable under'' relation.\label{tbl:translation}}
   \includegraphics{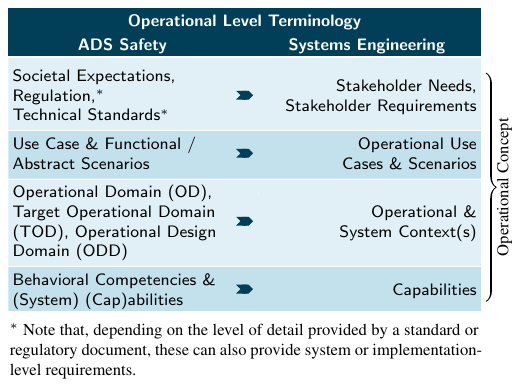}
   \vspace{-1.5em}
\end{table}

\subsubsection{Capturing Stakeholder Needs \& Requirements}
In Systems Engineering, the first step for defining an operational concept is the elicitation of stakeholder needs and the translation to actionable stakeholder requirements.
For traceable ADS (safety) assurance, this implies that sources of stakeholder needs (laws, standards, societal expectations, etc.) must be captured and represented in a structured way.
\cite{salem2024} gives examples of how such a structured representation of legal requirements for system behavior can be achieved.

\subsubsection{Defining Use-Cases \& System Context}
Following the elicitation of stakeholder needs and requirements, use cases are formulated that present a first abstract representation of \emph{what} the ADS should do.
Use cases subsume sets of scenarios in similar settings \cite{ulbrich2015}.

To detail use cases, so-called system or operational contexts are defined, which specify those entities with which the system must interact.
An (iterative) ODD definition determines, which entities can be part of a system context and which can, by that, be part of a use case and of a scenario.

Regarding AI-based systems, the system context is an important contribution to the input space definition: The elements of the system context are, e.g., sources of requirements for must-have class labels.

\subsubsection{Early Scenario-Based Safety Analyses and Behavior Specification}
The formulated use cases are typically detailed in scenarios.
System-level safety analyses and the definition of desired system behavior can already be conducted in compliance with ISO~26262 \cite{iso2018} or ISO~21448 \cite{iso21448} by using functional (or  abstract scenarios) without an immediate need to provide parameter ranges or choosing parameter values  \cite{salem2024}.

This enables the behavior-based formulation of safety goals and safety requirements, leaving open what kind of insufficiency actually causes a system-level hazardous event \cite{salem2024a} (i.e. E/E failures \cite{iso2018}, sensor-related performance insufficiencies \cite{iso21448}, AI-related performance insufficiencies \cite{iso8800} or specification insufficiencies \cite{iso21448}).
In other words, hazard and risk assessments can be performed on the behavior level. 
Safety-domain-specific analyses can follow, e.g. applying the failure mode models provided by ISO~26262, ISO~21448 or ISO/PAS~8800.

This concept follows the idea of ISO/PAS~8800, by explicitly separating between system-level and AI-specific analyses.
The resulting behavior specification provides an artifact that can be used to trace back additional AI-specific requirements, e.g. in the form of additional data labels that are required to comply with the behavior specification. 

\subsubsection{Definition of Behavioral Competencies}
The way that system capabilities or behavioral competencies have been described in the literature \cite{wasson2015, nolte2017, avsc2021}, they can provide a common starting point for decomposing system-level safety indicators into component-level performance metrics.
Capabilities are Systems Engineering concepts to capture required potentials of a system to achieve an outcome with a specified performance \cite{wasson2015}.
The AVSC consortium \cite{avsc2021} applies the concept of \emph{behavioral competencies} similarly.

Behavioral competencies are related to a behavior specification, as they provide a way of reasoning \emph{why} a system can show the required behavior.
As the concept is a combination of a desired potential (requirement), outcome (behavior), and performance, the definition of behavioral competencies can provide guidance for a traceable definition of pass-fail criteria from system-level safety indicators, as well as the definition of meaningful performance criteria for AI components, as discussed in \cite{troubitsyna2024}.

\section{Example Case Study}
To illustrate the concepts discussed in this section, we will provide a short, hypothetical case study that largely connects previously published work. 
Specified safety goals, behavioral competencies, and  requirements hence only serve illustrative purposes.
They are not reflecting existing or developed services or products by the involved industry partners.

\subsection{Stakeholder Needs}
For this case study, we assume stakeholder needs that require traffic-code-compliant behavior of the vehicle and a net risk that is ``lower'' than the risk caused by human drivers in comparable Operational Design Domains.

\subsection{Scenario and Context Definition}
To discuss the connection between system-level and AI-specific safety analyses, we consider a scenario that has been discussed in \cite{graubohm_assumptions_2023}.
Safety Goals and requirements at the behavioral level for this scenario have been specified in \cite{nolte2024}: An automated vehicle approaches a row of parked vehicles. Between two of the parked vehicles, a pedestrian is about to step onto the street and into the path of the automated vehicle.\footnote{A corresponding use case can be formulated as \emph{passing parked vehicles}, comprising a suite of similar scenarios.}
For the sake of the example, we assume that the ODD comprises straight inner-city roads with one lane in each driving direction and parking lanes parallel to these driving lanes.
Pedestrians and stationary vehicles of various vehicle types are also included in the ODD.

For this single scenario, the \emph{operational context} instantiates one pedestrian, a number of stationary vehicles, as well as the road with a parking lane as described above.

\subsubsection{Safety Analyses}
A \emph{hazard analysis} conducted at the behavior level can yield the following results:\footnote{For the case study, we omit a concrete risk assessment: exact numbers for assigned risks are irrelevant for showcasing the interplay between behavior-level and AI-specific analyses.}
\begin{itemize}
    \item A \emph{hazard} in the discussed scenario is the \emph{potential injury of vulnerable road users}.
    \item The discussed scenario becomes a \emph{hazardous scenario} in the SOTIF sense if the ego vehicle shows hazardous behavior in the given scenario.
    Such behavior can, e.g., be represented by a \emph{follow lane maneuver} at an \emph{inadequate speed} that prevents timely deceleration. Hence the maneuver can cause a collision with the pedestrian.
    \item A hazardous event that instantiates the hazard in the scenario is the \emph{collision of the ego vehicle with the pedestrian}.
\end{itemize}

A \emph{safety goal} that must be fulfilled in the scenario would be: \emph{Collisions with vulnerable road users must be prevented.}

\subsection{SOTIF-Specific Analysis Results}
Continuing the analysis in a SOTIF scope, it is possible to identify a \emph{functional insufficiency}: the ego vehicle is \emph{not able to correctly predict the behavior of occluded pedestrians}. 
The corresponding \emph{specification insufficiency} is the missing specification of \emph{occluded areas} which can contain pedestrians and which can occur in the given Operational Design Domain. 
The resulting triggering condition would be the \emph{presence of a pedestrian in an occluded area} in the given scenario.

Conducting these SOTIF analyses provides insights into how the risk of a collision with an occluded road user in the given scenario can be mitigated.
Risk mitigation can be achieved by a) including occlusions in the specification of a world model and b) defining according behavioral competencies for the automated driving system. The AVSC, e.g. defines the behavioral competency \emph{Responding to vulnerable road users (VRUs)} \cite[p. 9]{avsc2021}. Our scenario would extend this to \emph{Responding to \emph{occluded} vulnerable road users}.
In the process, these behavioral competencies would be specified by according (first functional and later technical) safety requirements.

\subsubsection{Connecting to AI-specific Risks}
Focusing on the connection to AI-specific risks, we can compare the discussion of the scenario with \cref{fig:general_uncertainty}.
In this example scenario the cause for the risk is an insufficient definition of the environment.
This is, at this time, independent of any specific system implementation.

If we assume that environment perception and decision making are performed by ML models, \cref{fig:ml_uncertainty} shows that AI-specific risks in the scenario stem from an insufficient definition of the input space for these ML models.
One possible risk mitigation measure that can be implemented in this context is the inclusion of occlusion scenarios (and possibly even the labeling of occlusions) in the datasets used for training the ML models.
In the sense of the ISO/PAS~8800 \cite{iso8800}, this results in \emph{dataset requirements} such as: \emph{The dataset for training the ML model must contain labeled occluded areas.}

\subsection{Discussion}
The hypothetical case study is just a rough textual example of single aspects regarding the transfer of traditional safety engineering approaches to AI safety.
All current standards emphasize the need for traceability between the specification, test results and safety performance indicators collected in the field.
To establish this kind of traceability, rich metamodels connecting behavior and ODD-specification will be required. 

First steps in this direction exist, e.g., with the A.U.T.O. ontology created in the VVMethods project \cite{westhofen2022} or our own previous work connecting Systems Engineering concepts with AD-Domain specific and behavior-related ontologies \cite{nolte2025, salem2024}.
However, a comprehensive approach that has been fully connected to AI-specific needs is still yet to be established.

Tooling that can reduce the effort of integrating model-based approaches into existing development processes will also be crucial: 
AI, function, and safety experts will need easily accessible interfaces to apply the ontologies and metamodels in their daily work without creating the burden of learning new description languages that do not target the core of their development activities.

%

\section{Conclusions \& Recommendations}
\label{sec:Conclusion}
This paper has emphasized the importance of identifying the reason for assurance challenges in AI-based automated driving systems.
We mapped out and highlighted the importance of differentiating which challenges are directly related to the nature of AI elements, and which are more related to the complexity of dealing safely with the open world that automated vehicle systems are deployed into.

The answer to the question in the title \enquote{\emph{What's \emph{really} different with AI?}} can be summarized along the lines of:
There are specific risks related to performance insufficiencies of AI-based system components (cf. ISO~8800 \cite{iso8800}).
However, principles of engineering rigor, which stem from traditional safety and systems engineering, are a necessary condition for building safe AI-based systems. 

In our view, this engineering rigor is rooted in diligent analyses for defining a system's operational concept. This can include, e.g., capturing Operational Design Domains, finding traceable models for ADS behavior in traffic, and defining the required behavioral competencies supported by scenario-based safety analyses.
These analyses can provide solid grounds for defining those system-level requirements and metrics as well as their decomposition to AI-related performance metrics, which is explicitly demanded by the current ISO~8800.
Again, the behavior-based approach outlined above is a \emph{starting point}.
The ADS safety assurance community has recently provided ideas and guidance \cite{salem2024, nolte2024, bsi1891, stettinger2024, barbier2024, fraade-blanar2025} contributing to such a behavior-based perspective.
Continuous efforts are required to further establish these concepts and consequently link them to AI assurance processes.

To summarize recommendations, we think that: a) It is crucial to assess, which benefits can be drawn from standards with broad AI definitions in what stage of the development process. b) A strong focus should be put on diligent problem space analyses, and on exploiting joint technology-neutral safety analyses as far as possible. This can provide an efficient, common starting point for SOTIF, Functional Safety, and AI safety analyses. c) Further research and standardization is needed for methods, ontologies and/or metamodels for the definition of behavior, as well as the definition of behavioral competencies. Finally, d), We would like to emphasize the conclusions by \citeauthor{troubitsyna2024} \cite{troubitsyna2024} that further research is required regarding the decomposition of system-level safety metrics to component-level and AI-related performance metrics.
%

\section*{Acknowledgement}
This paper is the product of ongoing discussions in the "Focus Field Safety and Risk" at the "German Round Table Autonomous Driving" initiated by the Federal Ministry on Digital and Transport.
We would like to thank Steffen Müller, Stefan Liening, and their teams at the BMDV DK20 for providing the frame for these discussions.
Further, we would like to thank the anonymous reviewers for their extremely constructive improvement suggestions, in particular regarding the case study and the reader guidance in section IV and the addition of a mini case study in section V.


\renewcommand*{\bibfont}{\footnotesize}

\printbibliography


\end{document}